\documentclass{article}

\usepackage{arxiv}

\usepackage[utf8]{inputenc} 
\usepackage[T1]{fontenc}    
\usepackage{hyperref}       
\usepackage{url}            
\usepackage{booktabs}       
\usepackage{amsfonts}       
\usepackage{amsmath}
\usepackage{nicefrac}       
\usepackage{microtype}      
\usepackage{lipsum}		
\usepackage{graphicx}
\usepackage{tikz} 
\usepackage{mdframed} 
\usepackage{svg}
\usepackage{subcaption}
\usepackage{float}
\usepackage{natbib}
\usepackage{doi}
\usepackage{xcolor}
\newcommand{\vect}[1]{\boldsymbol{#1}}

\title{Inverse Design of Copolymers Including Stoichiometry and Chain Architecture}

\author{
  Gabriel Vogel\\
  Department of Intelligent Systems\\
  Delft University of Technology\\
  Delft 2629 HZ, The Netherlands \\
  \texttt{g.vogel@tudelft.nl} \\
   \And
   Jana M. Weber\\
  Department of Intelligent Systems\\
  Delft University of Technology\\
  Delft 2629 HZ, The Netherlands \\
  \texttt{j.m.weber@tudelft.nl} \\
}

\hypersetup{
pdftitle={Inverse polymer design},
pdfsubject={q-bio.NC, q-bio.QM},
pdfauthor={Gabriel Vogel},
pdfkeywords={},
}

\begin{document}
\maketitle

\begin{abstract}
The demand for innovative synthetic polymers with improved properties is high, but their structural complexity and vast design space hinder rapid discovery. Machine learning-guided molecular design is a promising approach to accelerate polymer discovery. However, the scarcity of labeled polymer data and the complex hierarchical structure of synthetic polymers make generative design particularly challenging. We advance the current state-of-the-art approaches to generate not only repeating units, but monomer ensembles including their stoichiometry and chain architecture. We build upon a recent polymer representation that includes stoichiometries and chain architectures of monomer ensembles and develop a novel variational autoencoder (VAE) architecture encoding a graph and decoding a string. Using a semi-supervised setup, we enable the handling of partly labelled datasets which can be benefitial for domains with a small corpus of labelled data. Our model learns a continuous, well organized latent space (LS) that enables de-novo generation of copolymer structures including different monomer stoichiometries and chain architectures. In an inverse design case study, we demonstrate our model for in-silico discovery of novel conjugated copolymer photocatalysts for hydrogen production using optimization of the polymer's electron affinity and ionization potential in the latent space. 
\end{abstract}

\keywords{generative molecular design \and synthetic polymers \and higher-order information \and variational autoencoders \and transformers}

\section{Introduction}
Polymers have evolved into a cornerstone of modern society, finding applications across diverse domains such as food packaging~\citep{mangaraj2019polymerpackagingbiodeg}, textiles~\citep{morais2016textiles}, photovoltaics~\citep{liu2016PVpolymers}, clinical medicine~\citep{maitz2015polymersmedicine}, and many more. As the demand for innovative polymers surges, researchers face the task of navigating a vast design space. The space is uniquely characterized by the complexity of polymer materials which span from the chemical structure of monomers to diverse topologies and morphologies of the polymer material.
Traditionally, polymer design relies heavily on expert knowledge and trial-and-error of selected real-world experiments in the lab and in-silico simulations. Recently, machine learning (ML) has increasingly contributed to polymer informatics~\citep{yan2023rise, hatakeyama2023recent}, aiming to develop efficient strategies for polymer design. The core idea is to learn from existing polymer data, implicitly leveraging expert knowledge, to navigate the vast design space and find promising polymer candidates. These candidates can then be evaluated through in-silico experiments and synthesized and tested in the lab.

There are two prominent ML-based strategies to accelerate the discovery of new molecules~\citep{lavecchia2015MLDrugDiscovery, sanchez2018inverseMatterDesign, bilodeau2022generative4discovery, du2024GenAImolecular} also applicable to polymers. First, in direct design strategies~\citep{lavecchia2013virtual, gomez2016virtualscreeningmaterials}, ML models are trained to learn polymer structure-to-property relationships from known labelled polymer data. This approach can be used for virtual screening of polymer structure libraries to find the candidate with  the best predicted properties. This provides an efficient way to limit the number of experiments. However, this approach is constrained by a selection bias of predefined polymer structures to test, only focusing on a very small fraction of the full design space. In response, in inverse design strategies~\cite{sanchez2018inverseMatterDesign, sattari2021data} generative models are specifically trained to generate novel polymer structures with desired properties. One advantage of this approach is that candidate polymer structures do not need to be defined upfront. Labelled data and knowledge of the optimal property value is however still required. If trained correctly, this approach allows for iteratively optimizing toward novel polymers (not in the data library) with optimal properties (properties beyond the best values in the library). This can for instance be achieved through the combination of the generative model with optimization algorithms such as Bayesian optimization or genetic algorithms. In this work, we focus on the inverse design of polymers using generative modelling. 

\begin{figure}
    \centering
    \input{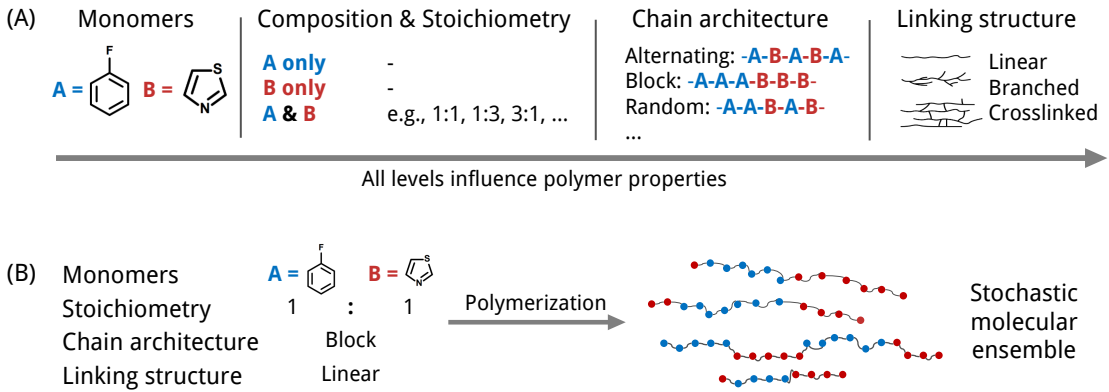}
    \caption{Two aspects of the complexity of synthetic polymer structures that need to be considered in the design due to their impact on the polymer properties. (A) Polymers possess a hierarchical structure from monomer structures, their composition (homopolymer, copolymer, etc.) and stoichiometry to chain architecture and linking structure. (B) Polymers are often stochastic materials composed of macromolecules of different lengths and weights.}
    \label{fig:PolymerStructure}
\end{figure}

An inverse design model requires a machine-readable representation that accurately captures the molecular structure and further can be generated by a generative model. This is especially challenging for polymers. Polymers possess a hierarchical structure spanning from the monomer chemistry to the polymer morphology. To a certain extent, all levels indicated in
Figure~\ref{fig:PolymerStructure}(A) impact the material's properties. At the smallest scale, polymers consist of monomer units with varying stoichiometries. Depending on the polymerization type and the reaction/processing conditions, the monomers can be arranged in diverse chain architectures, which are often described as polymer topologies. For instance, simply linear topologies with alternating, block, random, or other monomer arrangements exist. Other examples are lightly branched chains~\citep{zhu2011polymerbranching},  hyperbranched~\citep{hult1999hyperbranchedPolymers}, cyclic~\citep{hadjichristidis2001polymerscyclic} or star-shaped~\citep{bazan1991starPolymers} polymer topologies. Further, the processing conditions influence morphological traits such as crystallinity and therewith also the properties \citep{hatakeyama2023recent}. Besides, many polymers are stochastic. That means that they are ensembles of macromolecules of differing sizes and weights (see Figure~\ref{fig:PolymerStructure}(B)), which complicates the use of just one single polymer representation.


Numerous studies have employed descriptor-based fingerprinting in different variations to represent polymers as numerical vectors (see a comprehensive overview in~\citep{yan2023rise}). Polymer fingerprints have been widely used for predicting polymer properties~\citep{kuenneth2021polymer,kuenneth2022bioplasticdesign, doan2020predictionsGenome, pilania2019PHAfingerprintsPrediction, bhowmik2021prediction}, however and most notably, fingerprints are unsuitable for generative design. This is due to their non-invertibility, meaning that they cannot be uniquely mapped back and forth between the actual molecule and the fingerprint. In contrast, string and graph-based representations are invertible and thus appropriate for generative design.

Most string-based polymer representations, such as pSMILES~\cite{ma2020PI1m, kim2018polymerGenome}, build upon the SMILES~(Simplified Molecule Line Entry System) notation, adapting it for polymers by incorporating the repeating unit's SMILES and connection points using the * symbol. 
The pSMILES (or similar variants) has been used widely for the prediction of mechanical, thermal, thermodynamical, electronical and optical properties \citep{kuenneth2023polybert,xu2023transpolymer}. Moreover, it has been used to discover novel polymer structures with generative models~\cite{batra2020polymers,ma2020PI1m, ohno2023smipoly,kim2023OMG}. While pSMILES efficiently describes repeat unit chemistries, it falls short in depicting monomer combinations, higher-order structures, and polymer stochasticity. BigSMILES, an extension by \citet{lin2019bigsmiles}, offers greater flexibility, describing polymers as stochastic objects, accommodating multiple monomers and their connection points. However, BigSMILES lacks encoding for stoichiometry and processing history. G-BigSMILES \citep{schneider2024generative}, a recent work, extends the BigSMILES to include molecular weight distributions and connection probabilities, indirectly capturing monomer stoichiometry. The recent advances in the string-based polymer representations show promise for the development of more precise inverse design models, however these advanced representations have not yet been employed with generative models. 

Most polymer graph representations are simplified using only the molecular graph of one monomer or repeat unit \citep{park2022polymerGCN}. This can be applied for property prediction but also for generative design as shown in the work by \citet{liu2023graphGenDesign}. 
Recent works aim to move towards polymer graphs that represent macromolecules instead of only the repeat unit. One proposed approach known as PolyGrammar~\citep{guo2022polygrammar} makes use of hypergraphs to represent polymeric structures by explicitly defining sequences of monomers. It enables describing complex chain architectures and allows for generative design approaches, but does not account for the stochastic nature of polymer materials. Other approaches approximate the macromolecule by adding an additional edge (loop) between the connection points of the repeating unit \citep{antoniuk2022PolPeriodicGraph, gurnani2023MultitaskGNN}. Moreover, \citet{aldeghi2022graph} introduced a polymer graph that comprises multiple monomers as subgraphs, which are interconnected by additional weighted edges. The introduction of weighted edges enables the authors to capture the stochastic nature of polymer as monomer ensembles and facilitates the description of diverse polymer types (e.g., homo- and copolymers including stoichiometry) and various chain architectures (such as alternating, block, random, and graft structures).

Given the impact of structural and stochastic variability beyond the repeat unit on polymer properties, we believe that incorporating these factors is key to accurate inverse design of property-optimized polymers. Our work addresses this challenge. We build upon the recent graph representation~\citep{aldeghi2022graph} and use a Graph-2-string Variational Autoencoder~\citep{vogel2023graphtostring} for inverse design of copolymers. Our work generates polymers defined by their monomer chemistries, monomer stoichiometry, and chain architecture. 
We lead our generative model towards the design of optimal polymers through two steps. Firstly, we present a semi-supervised VAE setup to enable working with partly labelled and unlabelled data. Secondly, we employ two molecular optimisation strategies in the continuous latent space, which were previously demonstrated successfully in the literature~\citep{griffiths2020constrainedBO,sousa2021EvoinLS}. 
We illustrate our approach in a case study for the in-silico discovery of novel photocatalysts for green hydrogen production. 

\section{Methods}
This section first explains the employed graph and text-based polymer representations  briefly. Afterwards, we outline the general concept of VAEs and explain the detailed model architecture of the semi-supervised Graph-2-string VAE developed in this work. We then introduce the case study and dataset of copolymer photocatalysts for hydrogen production. Furthermore, we explain the evaluation metrics and conclude with details on the optimization in latent space.  
\subsection{Polymer representations}
\begin{figure}
    \centering
    \includegraphics[width=\textwidth]{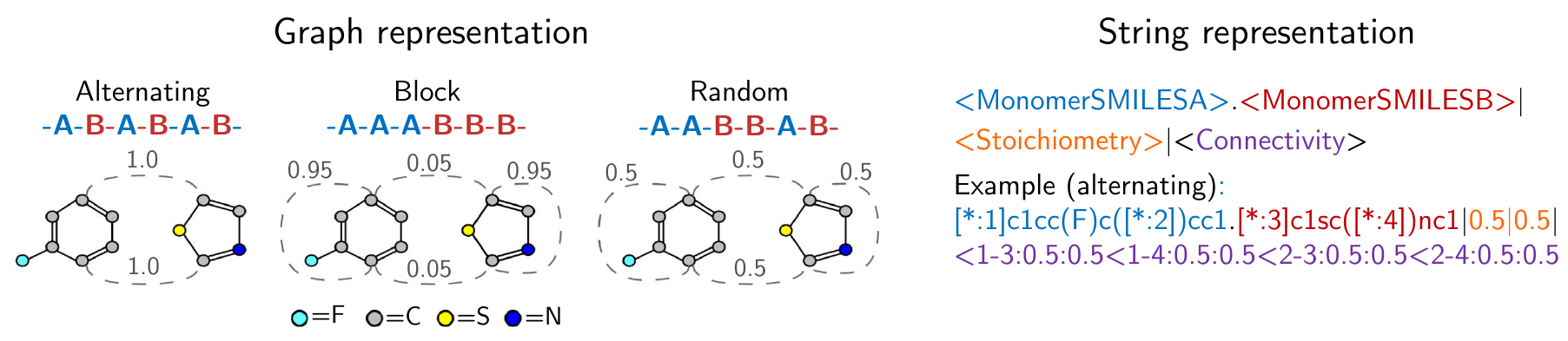}
    \caption{Polymer representations accounting for stoichiometry of monomer ensembles, the chain architecture and the stochastic nature of polymers. The graph representation is adopted from~\citet{aldeghi2022graph}, with stochastic edges (dashed) reflecting the connection propbabilities between monomers, i.e. reflecting the chain archtiecture. The string representation is a text-based description of the polymer graph representation, concatenating monomer SMILES, stoichiometry and connection probabilities.}
    \label{fig:Representation}
\end{figure}

Our approach integrates graph- and string-based representations, as depicted in Figure~\ref{fig:Representation}. Both represent the polymers as stochastic ensembles, including stoichiometry and chain architecture information.
\subsubsection{Graph representation}
Each polymer graph $G^{(i)}=\{V^{(i)},E^{(i)}\}$ consists of nodes (or vertices) $V_i$ representing atoms and edges $E_i$ representing bonds. In the graph representation by \citet{aldeghi2022graph}, the edges are weighted according to the probability that they occur in the polymer. Bonds within the monomer structure have the weight $w=1.0$ and bonds between monomers have a weight $w\in(0,\,1.0]$. The weights reflect how monomers are connected, essentially depicting the chain architecture of the polymer. Furthermore, the stoichiometry is incorporated as node weights, where nodes of the same monomer have the same weight. The node weights are used during pooling in the graph neural network (GNN) introduced by \citet{aldeghi2022graph}. We make use of their GNN architecture as encoder block for the VAE, see Section~\ref{sec:modelarchitecture}. 
\subsubsection{String representation}
The string representation encapsulates stoichiometry and connection probabilities as numerical values next to the monomer SMILES. A polymer string $\vect{x_S}^{(i)}$ can be formally described as a sequence of tokens $\vect{x_S}^{(i)}=\{\vect{x_S}_1^{(i)},\vect{x_S}_2^{(i)},...,\vect{x_S}_N^{(i)}\}$, where $\vect{x_S}_i^{(i)}$ can be a SMILES token describing the monomer chemistry or a numerical token as part of the stochiometry or chain architecture.  
We use a tokenization scheme that combines a SMILES tokenizer and numerical number tokenizer adapted from the Regression Transformer~\cite{born2022RT}. In Appendix~\ref{app:tokenization} we show an example tokenization of a polymer string. 

\subsection{Model}\label{sec:modelarchitecture}

We build on the general framework of a variational autoencoder (VAE)~\citep{kingma2013VAE}, a probabilistic generative model. This model consists of an encoder network $q_{\vect{\phi}}(\vect{z}|\vect{\vect{x}})$ that maps high-dimensional data $\vect{x}$ to a latent space distribution $\vect{z}$, which is approximated by a Gaussian distribution with mean $\mu$ and variance $\sigma^2$. Additionally, it includes a decoder network $p_{\vect{\theta}}(\vect{x}|\vect{z})$ that maps samples from the latent space $\vect{z}$ back to the data space, aiming to reconstruct the original input $\vect{x}$. 

\subsubsection{Graph-2-string variational autoencoder}
\begin{figure}
    \centering
    \includegraphics[width=\textwidth]{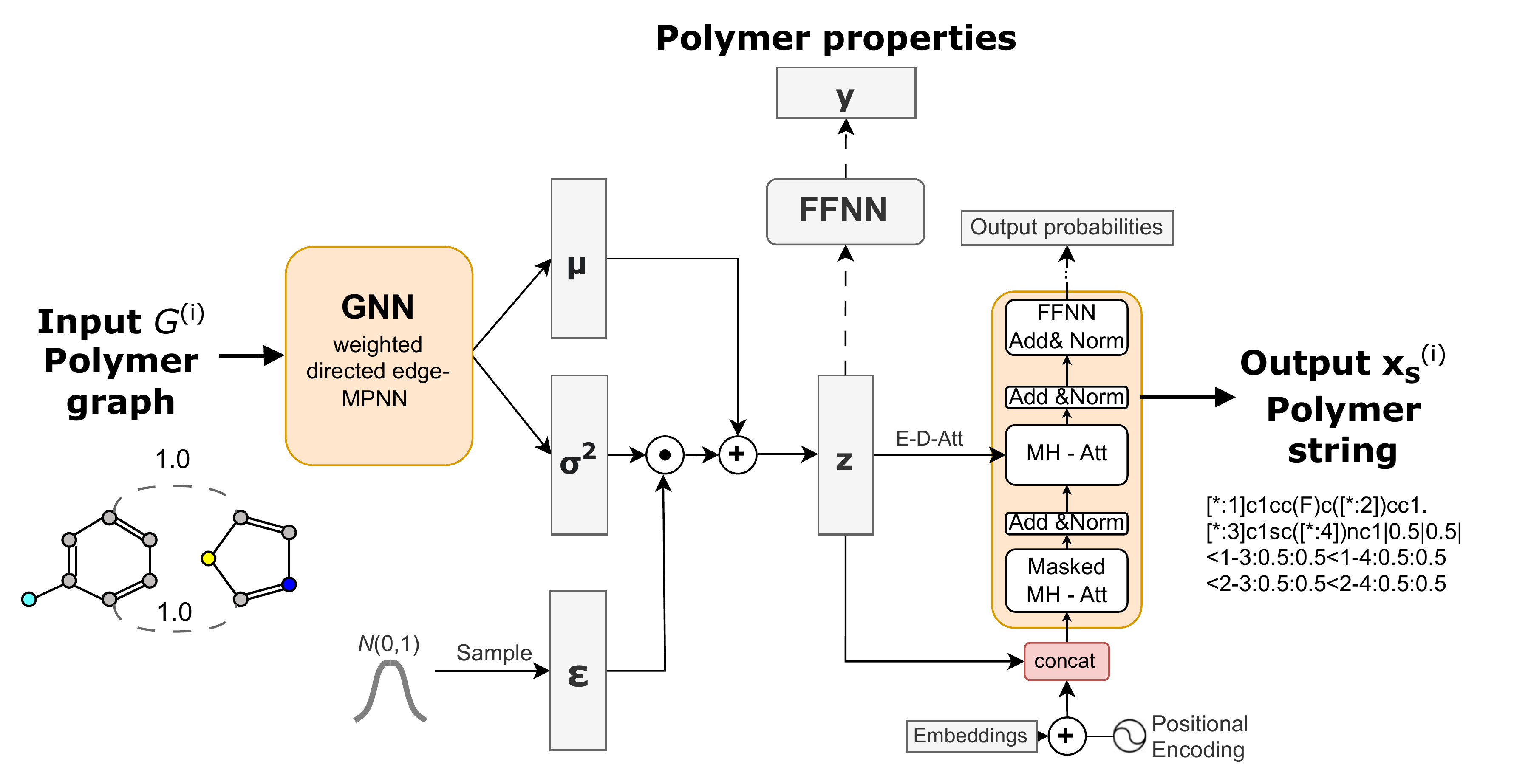}
    \caption{Semi-supervised Graph-2-string VAE architecture. The polymers are represented as graphs and encoded in a wD-MPNN (weighted directed message passing neural network) to obtain mean $\mu$ and variance $\sigma^2$ tensors. The latent representation $\vect{z}$ is sampled from a normal distribution parametrized by $\mu$ and $\sigma$, using the reparametrization trick. The latent representation $\vect{z}$ is fed both to a feed forward neural network to predict polymer properties (for labelled data) and to the Transformer decoder to reconstruct the polymer in string format. }
    \label{fig:ModelArchitecture}
\end{figure}

In the graph-2-string VAE we encode polymer graphs $G_i$ to a latent embedding $\vect{z}$ and use it to predict polymer properties and decode an equivalent polymer string $\vect{x_S}^{(i)}$ instead of the graph (Figure~\ref{fig:ModelArchitecture}). \citet{dollar2023efficient} also demonstrated the benefits of using a graph-to-string VAE for small molecular design. Very recently, another work developed a graph-to-string VAE architecture in the field of drug design~\citep{muller24combining}, combining graph attention neural networks and recurrent neural networks.

To encode a polymer graph we re-implement the proposed weighted directed edge message passing neural network (wD-MPNN)~\citep{aldeghi2022graph}. We use one wd-MPNN block with 3 layers and subsequently two parallel blocks that output a mean $\mu$ and variance $\sigma^2$ vector, respectively~\citep{vogel2023graphtostring}. Then we use the reparametrization trick~\citep{kingma2013VAE} to obtain the latent embedding $\vect{z}$. 

The latent embedding is fed to a Transformer-based~\cite{vaswani2017attention} language model consisting of four sequential layers with each four attention heads to decode the equivalent polymer string. In addition to the encoder-decoder cross attention between $\vect{z}$ and previously generated token embeddings, we concatenate $\vect{z}$ with each token embedding after the positional encoding, slightly different to \citet{fang2021transformer} who added it element wise. This was motivated by work in the natural language domain showing that the combination of VAEs and Transformer decoder requires modifications that regulate how the latent space is fed to the decoder \citep{fang2021transformer, li2020optimus}. This modification improved the reconstruction ability of the model substantially as demonstrated in \cite{vogel2023graphtostring}. 

We further include knowledge about the properties in our model to generate a property-organised latent space. We adopt the approach of \citet{gomez2018automatic} that adds an additional feed-forward neural network taking the latent embedding $\vect{z}$ as input to predict a property $y$ (or multiple) that is jointly trained with the VAE architecture consisting of encoder and decoder networks.

\subsubsection{Loss function}
The loss in Equation~\eqref{eq:LossSupervisedVAE} to train the graph-2-string VAE architecture consists of a reconstruction loss term $\mathcal{L}_{Rec}$, Kullback-Leibler divergence loss $\mathcal{L}_{KLD}$ and an additional property loss term $\mathcal{L}_{y}$. For additional details on the deriviation of the loss see Appendix~\ref{app:lossderivation}. We further use two hyperparameters $\beta$ and $\alpha$ to balance the loss terms. 
\begin{equation}\label{eq:LossSupervisedVAE}
    \mathcal{L}=\mathcal{L}_{Rec} + \beta \cdot\mathcal{L}_{KLD} + \alpha\cdot \mathcal{L}_{y}
\end{equation}
The reconstruction loss $\mathcal{L}_{Rec}$ is calculated as the weighted cross entropy loss between the ground truth and predicted polymer string given the latent representation $\vect{z}$. Further, as will be outlined in Section~\ref{sec:data}, we have a combination of labelled and unlabelled data. To handle partially labelled data, we introduce a mask $m$ for the property loss 
\begin{equation}\label{eq:LossSemiSupVAE}
    \mathcal{L}=\mathcal{L}_{Rec} + \beta \cdot\mathcal{L}_{KLD} + \alpha\cdot m\cdot \mathcal{L}_{y},
\end{equation} 
ensuring it is calculated solely for labelled data. This approach limits gradient computation to labelled instances, thereby updating the property prediction network based exclusively on labelled data in a batch with $N$ samples. Given two ($M=2$) continuous properties $P=\{p_1, p_2\}$ of interest (see Section~\ref{sec:data}), we design the neural network to output two property values and calculate the masked property prediction loss as the mean squared error averaged over the two properties and the number of samples in the batch.
\begin{equation}
m\cdot \mathcal{L}_{y} = \frac{1}{N} \sum_{i=1}^{N} \left( \frac{1}{M} \sum_{j=1}^{M} \text{mask}(i,p_j) \cdot (y_{p_j, i} - \hat{y}_{p_j, i}))^2 \right)
\end{equation}


\subsection{Case study and dataset}\label{sec:data}
We test our model in a case study to design novel conjugated copolymers which are emerging as promising organic photocatalysts for green hydrogen production through photocatalytic water splitting. The process involves a photocatalyst that absorbs light to generate charge carriers which reduce protons to hydrogen while oxidizing water or an electron donor. Addressing the vast synthetic diversity of conjugated polymers, \citet{bai2019accelerated} conducted a high-throughput screening study to explore various copolymers, analyzing their hydrogen evolution rate (HER) as a measure of photocatalytic activity. Ultimately, we aim to find candidate polymers with improved catalytic activity (see Sections~\ref{sec:propOptMethods} and~\ref{sec:inversedesign}). 

\begin{figure}[H]
    \centering
    \includegraphics[width=0.7\textwidth]{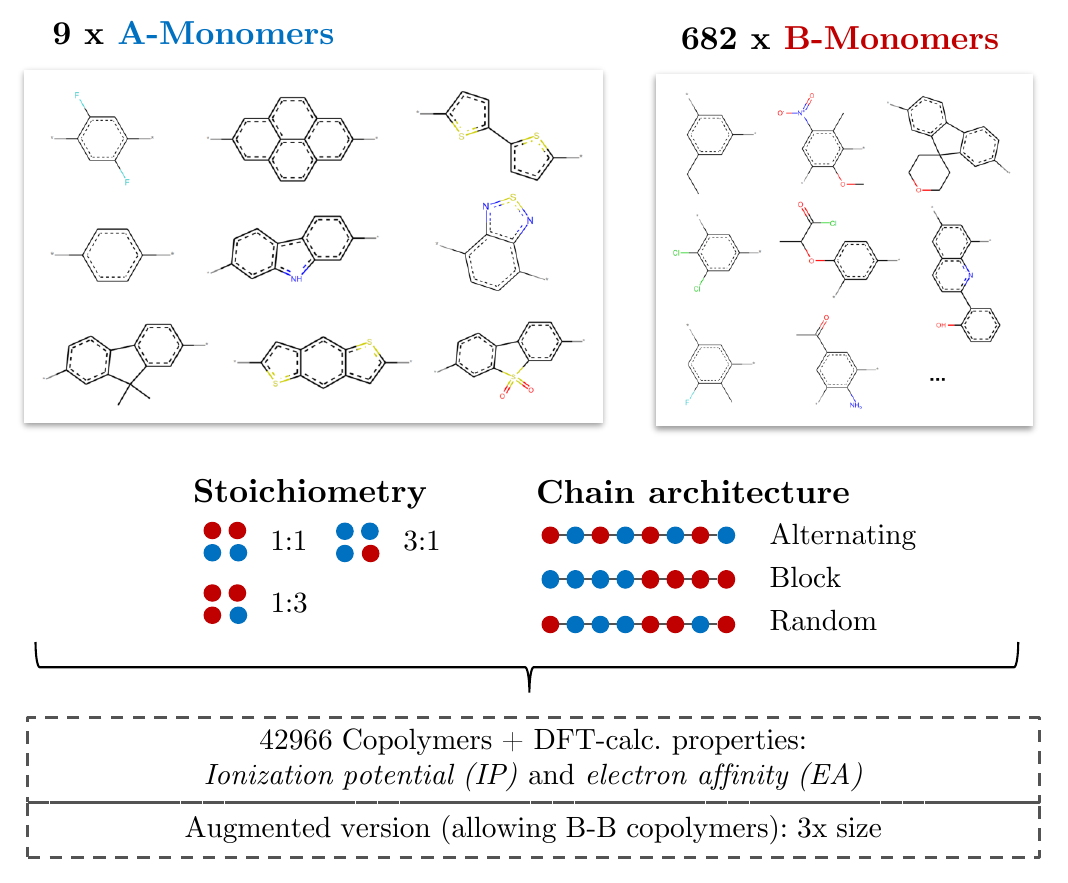}
    \caption{Copolymer photocatalyst dataset from \cite{aldeghi2022graph} that is used in this work. The polymer space consists of 9 A-monomers and 682 B-monomers that are combined in three stoichiometries (1:1, 1:3, 3:1) and three chain architectures (alternating, block, random). This forms a dataset of 42966 copolymers, including DFT-calculated polymer properties ionization potential (IP) and electron affinity (EA). We create an augmented data set without the property labels (ca. 3 times the size) by allowing the combination of B-B copolymers.}
    \label{fig:dataset}
\end{figure}

We base our study on the dataset introduced by \citet{aldeghi2022graph} which is built upon the conjugated copolymer space of \citet{bai2019accelerated}. As shown in Figure~\ref{fig:dataset}, the dataset combines eight A-monomers with 682 B-monomers in stoichiometries of 1:1, 1:3, and 3:1 and three types of chain architectures (alternating, random, block). This leads to 42966 copolymers. The dataset additionally reports the ionization potential (IP) and electron affinity (EA), determined with DFT calculations for all 42966 polymer candidates \citep{aldeghi2022graph}.
Additionally, we construct an augmented version of the data set, allowing for B-B copolymers to increase the diversity of the chemical space of monomer A, comprising 138447 copolymers (ca. 3x the size of original dataset). The labels IP and EA are only available for the original data points, hence, we obtain a partly labelled dataset. 

\subsection{Training and evaluation}\label{sec:evaluation}
We split the dataset in 80\% training data, and 10\% validation and 10\% test data. Since we have multiple data points per monomer combination (varying stoichiomtry and chain architecture), we split the data by monomer combinations to prevent data leakage, i.e. ensuring that there are no monomer combinations in the test- or validation set that also occur in the training set. During training we apply the default teacher forcing in the Transformer decoder and use early stopping based on the evaluation on the held out validation set. During inference, i.e. the novel generation of molecules from latent embeddings $\vect{z}$, we use beam search with a beam size of five to decode the polymer strings. This maximizes the final probability of the sequence and helps with generating valid SMILES. Where applicable, we adopted the hyperparameters of~\citep{vogel2023graphtostring}, which is the unsupervised version (only encoder and decoder) of the model used in this work. To ensure a more stable training we use a cyclical schedule for $\beta$~\citep{fu2019cyclical}. Additionally, we performed a hyperparameter search over $\beta$ and $\alpha$ to balance the contribution of the loss terms in Equation~\eqref{eq:LossSemiSupVAE}. Note, that we limited the hyperparameter search of these parameters to a fixed number of evaluations in a random search approach and subsequent fine-grained grid search.

During the hyperparameter search and to assess the final model performance, we calculate quantitative metrics that have been established in the small molecule domain: reconstruction, validity, novelty, and uniqueness, defined in Table~\ref{tab:Metrics}. Reconstruction is evaluated on the test set, encoding each sample to a latent point $\vect{z}=\vect{\mu}$ and passing it to the decoder to produce the polymer string representation. Novelty and Uniqueness are evaluated for the sampled set which consists of 16000 polymers randomly sampled from Gaussian noise. Validity is calculated both for the reconstructed test set as well as the sampled set. In the results in Section~\ref{sec:resultsDiscussion} we use the best performing model according to these model metrics and qualitative assessments of the latent space structure. In Section~\ref{sec:ModelSelection} we explain in more detail how we combine the quantitative and qualitative assessment.

\begin{table}[H]
  \caption{Common quantitative metrics used for evaluation of different models during hyperparameter search. Reconstruction is evaluated for the test set while novelty and uniqueness are evaluated for a sampled set of 16000 polymers sampled from Gaussian noise. }
  \label{tab:Metrics}
  \centering
  \begin{tabular}{p{0.15\textwidth}p{0.2\textwidth}p{0.55\textwidth}}
    \toprule
    Metric     & Data     & Definition \\
    \midrule
    Reconstruction  & test set  & Percent of correctly reconstructed molecules    \\
    Validity     & test set, sampled set &  Percent of valid molecules evaluated using RDKit's molecular structure parser \cite{bilodeau2022generative4discovery}     \\
    Novelty     & sampled set       & Percent of molecules not present in training set \cite{bilodeau2022generative4discovery}  \\
    Uniqueness     & sampled set       & Percent of unique molecules \cite{bilodeau2022generative4discovery}\\
    \bottomrule
  \end{tabular}
\end{table}

\subsection{Property optimization and inverse design}\label{sec:propOptMethods}
To generate molecules with desired properties, we employ optimization in the numerical latent space of our trained VAE model. We define the 32 numerical latent dimensions as input variables for an optimization algorithm, denoted as $\vect{z}=\{z_1, z_2, ..., z_{32}\}$.
The objective function is composed of target values for the predicted copolymer properties from the latent space. The ideal scenario would be to maximize for the HER, however, this property is not reported for the dataset we used to train our model. We use the observation~\citep{bai2019accelerated}, that the two electronic properties IP and EA are correlated to the HER of copolymer catalysts. Optimal target values for IP and EA can be derived for instance using expert knowledge, targeting a specific chemical region that is known to perform well (see Appendix~\ref{app:DesignWOPPO} and~\ref{app:differentObjFuns}), or through observed trends in high-throughput studies. Hereinafter, we used the last option, deriving target values of EA and IP from the analysis in~\citet{bai2019accelerated}.

Based on the findings in \citep{bai2019accelerated}, we determine that IP$\approx1$ and more negative values of EA are indicative of better HERs and thus correlate to better photocatalysts. In Appendix~\ref{app:choiceTargetVals} we elaborate on the choice of the exact target values of IP and EA. Consequently, we seek to minimize the overall objective function  $f(\vect{z})$
\begin{equation}
    f(\vect{z}) = |pp_{IP}(\vect{z})-1| + pp_{EA}(\vect{z}) + pen_{inv}(\vect{x_S}).
    \label{eq:propOpt}
\end{equation}
Here, $pp_{IP}(\vect{z})$ and $pp_{EA}(\vect{z})$ represent properties predicted from latent inputs using the trained property prediction neural network of our model. The last term assigns a penalty if the decoded string $\vect{x_S}=p_{\vect{\theta}}(\vect{x}|\vect{z})$ is invalid. To perform the optimization task in latent space, common approaches are gradient-based optimization, genetic algorithms (GA) or Bayesian optimization (BO). In this work we focus on BO and GAs for the optimization task.



\subsubsection{Bayesian optimization}
We implement BO using the Bayesian optimization Python package~\citep{BayesOptPackage}. We run the task as single objective optimization with the two objectives aggregated in one objective function, as defined in Equation~\eqref{eq:propOpt}. The BO first fits a surrogate model, here a Gaussian Process, to represent the mapping between input variables and the objective function for observed data points including a quantification of the uncertainty of unobserved areas. Then, iteratively the algorithm uses an acquisition function to select new evaluation points. As acquisition function, we employ the upper confidence bounds method that takes as input the expected objective $\mu(\vect{z})$ and uncertainty $\sigma(\vect{z})$ given by the Gaussian process as function of the input variables $\vect{z}$
\begin{equation}
    UCB(\vect{z};\kappa)=\mu(\vect{z})+\kappa \sigma(\vect{z}),
\end{equation}
where $\kappa$ balances exploration and exploitation (higher $\kappa$ favors more exploration). We use the default value for $\kappa$ of 2.576. \\
A substantial number of works discuss good practices~\citep{siivola2021BOgoodpractices} or propose strategies to overcome common challenges of Bayesian optimization particularly in the relatively high-dimensional latent spaces of VAEs, e.g.,~\citep{griffiths2020constrainedBO,maus2022localBO,tripp2020sampleeffBO}. A general issue encountered while performing any optimization in the latent space of VAEs is the presence of holes or low-confidence regions, which do not align with the high-confidence regions produced by the encoder. The optimization algorithm often tends to exploit these regions to optimize the objective, leading to a mismatch of the predicted and the real properties of the decoded polymer.
To address this, we designed an approach aimed at ensuring high confidence in property predictions $pp_{IP}(\vect{z})$ and $pp_{EA}(\vect{z})$. This involves correcting points sampled by BO, denoted as $\vect{z_{\text{BO}}}$, by decoding and re-encoding them to obtain $\vect{z'}\approx\vect{z_{\text{BO}}}$. Consequently, we ensure that the point lies within a high-confidence region of the latent space, suitable as input to the property predictor network. Thus, we achieve more accurate property predictions (see also Appendix~\ref{app:propPred}), which mitigates discrepancies between the predicted properties used for objective function evaluations and the corresponding decoded molecules.

\subsubsection{Genetic algorithm}
For the implementation of the GA we use the multi-objective Non-dominated Sorting Genetic Algorithm (NSGA-II)~\citep{NSGA2} from the pymoo python package~\citep{pymoo}. We use Latin Hypercube Sampling to ensure that the initial population covers the entire latent space uniformly. By iteratively performing selection, cross-over and mutation of the samples in a population, the GA evolves the population towards optimal or near-optimal solutions over successive generations. As crossover we use Simulated Binary Crossover (SBX), a method to combine previous solutions (parents) and produce offsprings. For each latent dimension, the parents' values are averaged and perturbed (small changes) according to a perturbation factor. It preserves useful properties of the parents and effictively balances exploration and exploitation. Lastly, we use Polynomial Mutation (PM), commonly paired with SBX, which performs additional slight perturbations of the offspring's latent dimensions generated by SBX. This helps to avoid local optima and enhance the search process. A mutation probability and perturbation factor control the probability for each latent dimensions to be perturbed and the extent of change, respectively. For both SBX and PM we use the default settings of the python package.\\
To tackle the problem of low-confidence regions in the latent space we implement a repair mechanism after the offsprings (new individuals after crossover and mutation) have been reproduced. We adjust these offsprings $\vect{z_{\text{offspring}}}$, by decoding and re-encoding them to obtain $\vect{z'_{\text{repaired}}}\approx\vect{z_{\text{offspring}}}$ to ensure they are within high-density regions of the latent space. These points are then evaluated to calculate the objectives. 

\section{Results and Discussion}\label{sec:resultsDiscussion}
In this section we demonstrate the generative capabilities of our model to generate novel copolymers including their stoichiometry and chain architecture. We first assess the overall model performance as a generative model in terms of reconstruction and validity, novelty, and uniqueness of sampled polymers. Further, we analyse the structure and smoothness of the latent space qualitatively. Then we test our model for inverse design of property-optimized organic copoloymer photocatalysts for improved green hydrogen production.
 
\subsection{Model evaluation}\label{sec:modelEvaluation}
\begin{figure}
    \centering
    \input{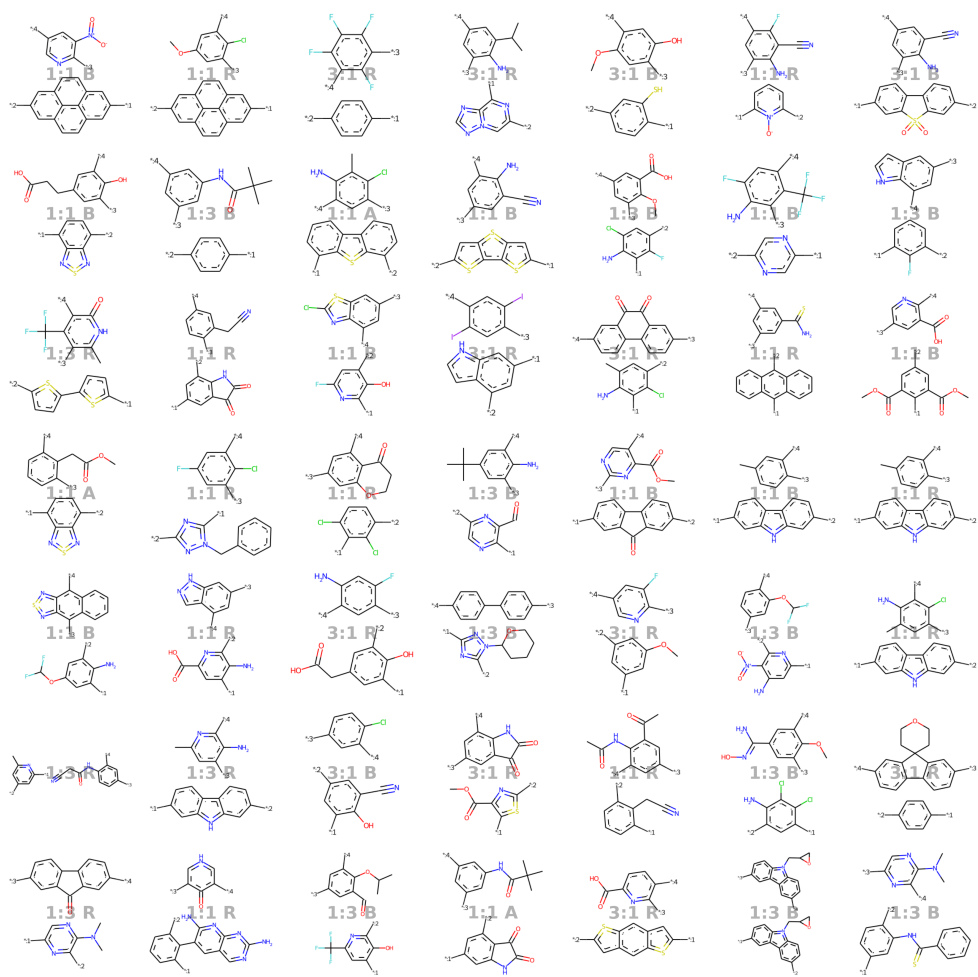} 
    \caption{56 example copolymers sampled from Gaussian noise. All sampled polymers belong to the class of conjugated copolymers as the training data, displaying a wide variety of monomer structures combined in different chain architectures and stoichiometries.}
    \label{fig:moleculesRandom}
\end{figure}

Our model demonstrates good performance in generating novel, diverse, and chemically valid polymers and reconstructing complex copolymer structures. With the best hyperparameter configuration (see Section~\ref{sec:ModelSelection}) we achieve a reconstruction accuracy of 68\% for copolymers in the testset, including their higher-order structural information (chain architecture and stoichiometry). Further, the model generates novel (81\%) and unique (98\%) polymers. The reconstructed test set exhibits 99\% valid and the generated sample set 96\% chemically valid molecules. We observe that the overall novelty (81\%) is composed of novel monomer chemistries (25\%), but also a substantial fraction of novel unseen combinations of monomers from the training set (81\%-25\% = 56\%). Further, the sampled molecules cover all combinations of stoichiometries and chain architectures from the dataset. The results demonstrate the generative capabilities, but also show room for improvement in terms of monomer novelty and possibly novelty on other structural levels (stoichiometry and chain architecture) through a greater dataset diversity.
Figure~\ref{fig:moleculesRandom} presents 56 example polymers sampled randomly from the latent space  after model training. The generated copolymers display a wide range of structures, showcasing various conjugated copolymers with differing monomer chemistries, stoichiometries, and chain architectures.

\subsection{Latent space}
In the following section, we demonstrate that the latent space (LS) is smooth and organized in a chemically meaningful way. These characteristics are of great importance when sampling novel polymers and using the latent space for property-guided polymer discovery. 
\subsubsection{Latent space organization}\label{sec:LSstructure}
\begin{figure}
    \centering
    \includegraphics[width=0.6\textwidth]{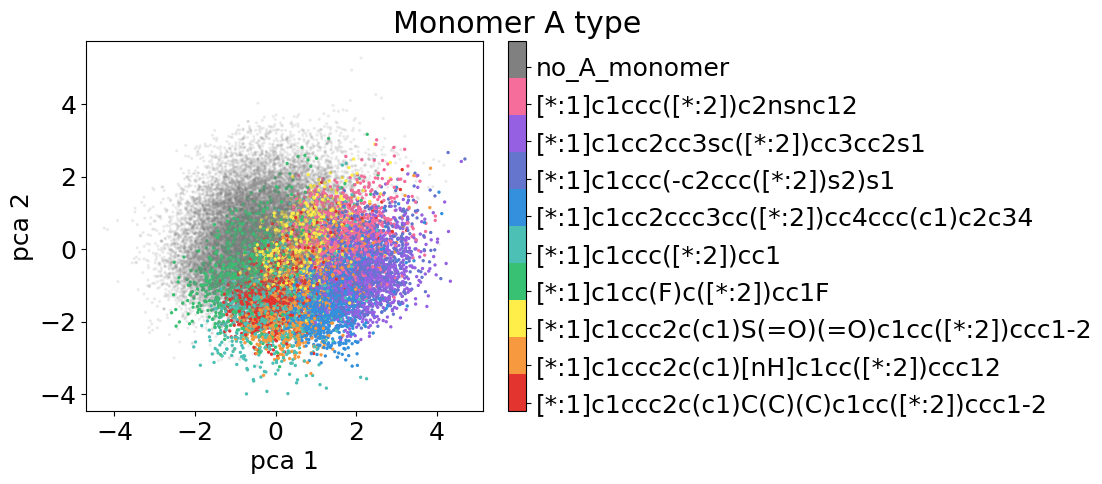}
        \caption{Visualization of the two first principal components of the latent space of the training data. Coloring by A monomer type reveals organization according to the monomer chemistry.}
        \label{fig:LSAMon}
\end{figure}

\begin{figure}
    \centering
    \begin{subfigure}[b]{0.35\textwidth}
        \centering
        \includegraphics[width=\textwidth]{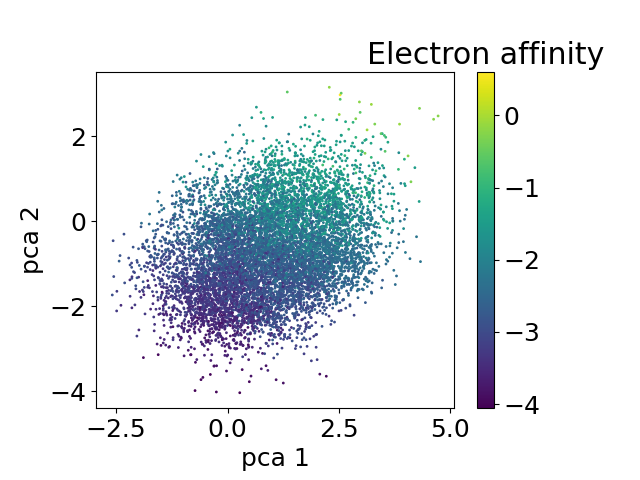}
        \caption{Colored by EA}
        \label{fig:LSIP}
    \end{subfigure}
    \begin{subfigure}[b]{0.35\textwidth}
        \centering
        \includegraphics[width=\textwidth]{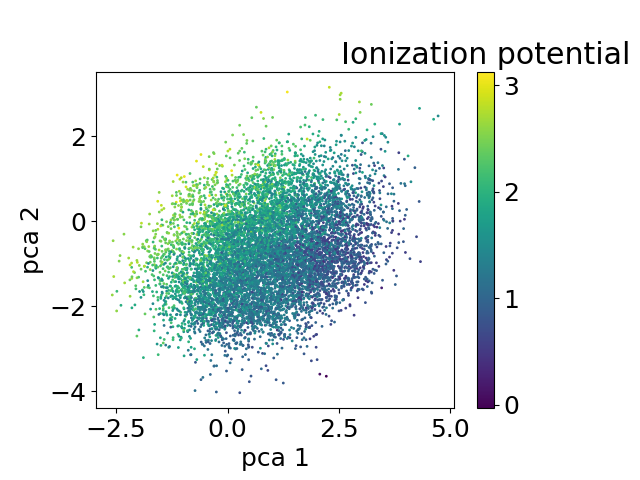}
        \caption{Colored by IP}
        \label{fig:LSEA}
    \end{subfigure}
    \caption{Visualization of the two first principal components of the latent space of the training data. Coloring according to polymer properties reveals that the latent space shows property gradients.}
    \label{fig:LatentSpaceplotsProperties}
\end{figure}

Figure~\ref{fig:LSAMon} visualizes the first two principal components of the training dataset encoded to the LS. We observe that the LS of the best model primarily organizes according to monomer chemistry. Figure~\ref{fig:LatentSpaceplotsStoichCon} in the Appendix shows that the LS is more locally structured according to stoichiometry and chain architecture. This organization aligns with the intuitive understanding that polymers sharing the same monomer types are inherently more similar, regardless of differences in chain architecture. Further, we verify that the LS structures according to the ionization potential (IP) and electron affinity (EA), which facilitates successful optimization in the LS (see Section~\ref{sec:inversedesign}). Figure~\ref{fig:LatentSpaceplotsProperties} illustrates distribution of property values and gradients in the PCA plot of the latent space, demonstrating the model's ability to capture property-structure relationships. 
\subsubsection{Smoothness of latent space}
\begin{figure}
    \centering
    \input{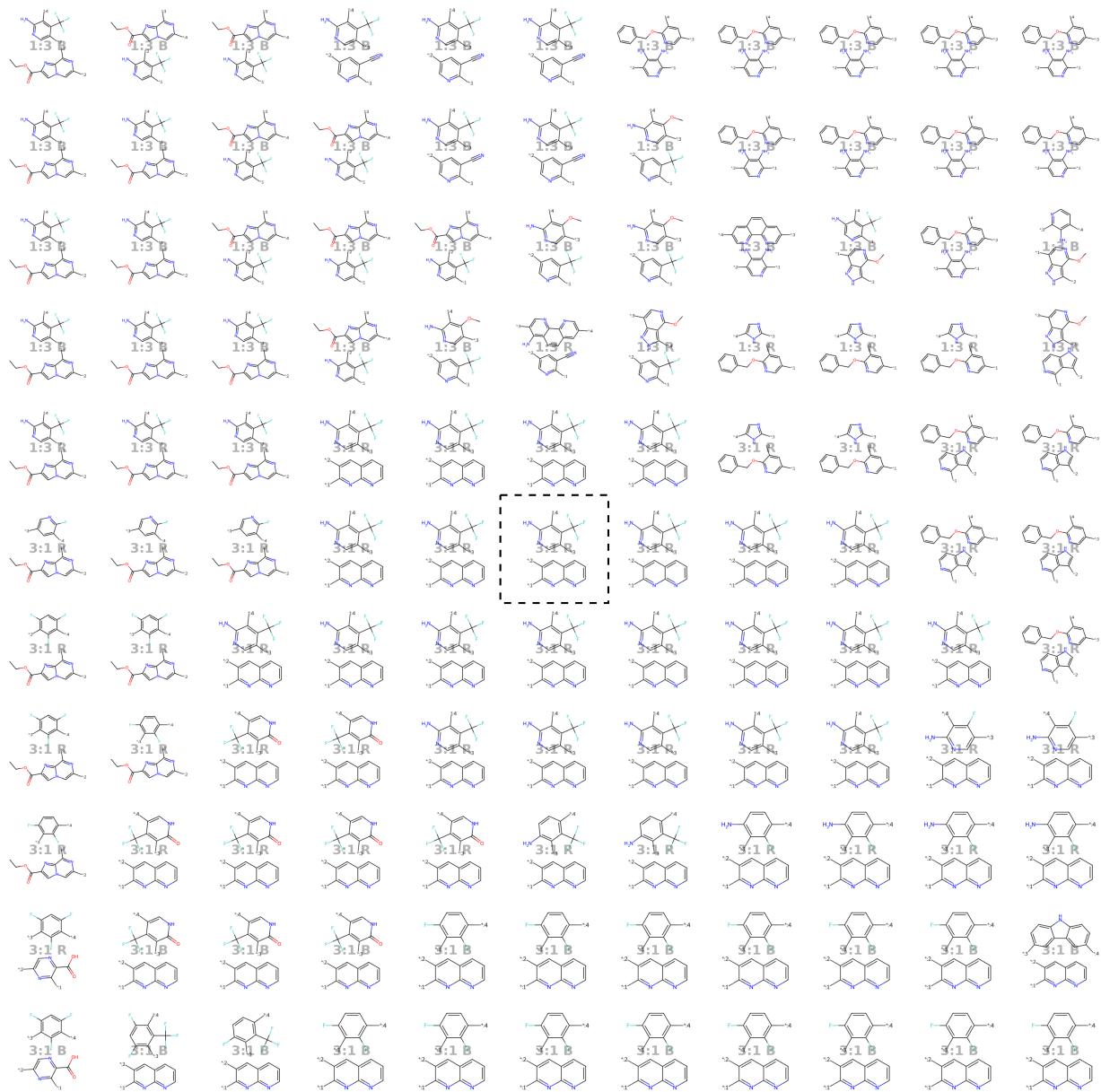} 
    \caption{Visualization of the molecular neighborhood using two orthogonal vectors in latent space as proposed by \citet{kusner2017grammarVAE}. Moving around the random seed molecule in black dashed lines in the center, we can observe step-wise changes in monomer A (bottom one), monomer B (top one), stoichioemtry and chain architecture.}
    \label{fig:moleculesNeighborhood}
\end{figure}

In line with the approach described by \citet{kusner2017grammarVAE}, we visualize the neighborhood of a seed molecule in Figure~\ref{fig:moleculesNeighborhood} to inspect the smoothness of the latent space. We start by defining two random orthogonal unit vectors (of dimension 32) in the latent space, scaled by the variance of a sample batch of molecules. From the seed molecule (in black dashed lines), we then explore its local neighborhood on a grid defined by these vectors. Each copolymer in Figure~\ref{fig:moleculesNeighborhood} comprises an A monomer (bottom) and a B monomer (top). We observe that monomer A remains unchanged for more steps than monomer B. Also, monomer B exhibits more gradual changes in the latent space, often altering only one atom type, a side chain, or functional group. This discrepancy is attributable to the different variety of A and B monomers in the training data: there are fewer A monomer chemistries compared to B monomer chemistries, resulting in a smoother latent space for monomer B. Note that the stoichiometry and chain architecture remain the same in the direct neighborhood of the seed molecule. This consistency is expected for stoichiometry, as changing from 1:1 to either 1:3 or 3:1 would significantly alter the nature of the polymer and thus should not be modified in the immediate neighborhood. For the chain architecture, the impact is slightly less pronounced, but changing from an alternating to a block structure would still substantially alter the polymer's nature. 

\subsection{Hyperparameter selection}\label{sec:ModelSelection}
As detailed in Section~\ref{sec:evaluation}, to select the best model we conducted a hyperparameter search over the weights $\beta$ and $\alpha$ in the loss terms in Equation~\eqref{eq:LossSemiSupVAE} and evaluated the model using the performance metrics from Table~\ref{tab:Metrics}. 
We observed that lower values of  $\beta$ and $\alpha$ enhance reconstruction accuracy by prioritizing the reconstruction term in the loss function. However, achieving high novelty and validity in randomly generated samples requires larger $\beta$ values. If $\beta$ is too small, the latent space becomes discontinuous with "holes" or regions of low confidence, leading to invalid molecules when sampled. We found that for this dataset, $\beta$ values between 0.0003 and 0.0004 work best to optimize the trade-off between reconstruction, validity and novelty. 
Additionally, we found that a well-structured latent space (LS) is crucial for successful inverse design using optimization algorithms. Ideally, the LS should organize polymers by their properties, as demonstrated in Section~\ref{sec:LSstructure}. Increasing the weighting factor $\alpha$ improved the organization of molecules regarding their properties, which was visually confirmed by examining PCA plots of the latent space showing clear property gradients (Figure~\ref{fig:LatentSpaceplotsProperties}). A sufficiently high $\alpha$ also leads to accurate property predictions, as verified in Appendix~\ref{app:propPred}. Contrary, increasing $\alpha$ too much harms the reconstruction performance.
As a result, we select the best model ($\alpha=0.2$ and $\beta=0.0004$) according to both the quantitative metrics and visual inspection of gradients in the latent space. In the subsequent section, we use this model for the inverse design approach. Note, that the hyperparameter tuning does not guarantee reaching a globally optimal configuration for the inverse design problem and may vary for new datasets. 

\subsection{Inverse design of novel polymer photocatalysts for hydrogen production}\label{sec:inversedesign}
\begin{figure}
    \centering
    \begin{subfigure}[b]{0.45\textwidth}
        \centering
        \includegraphics[width=\textwidth]{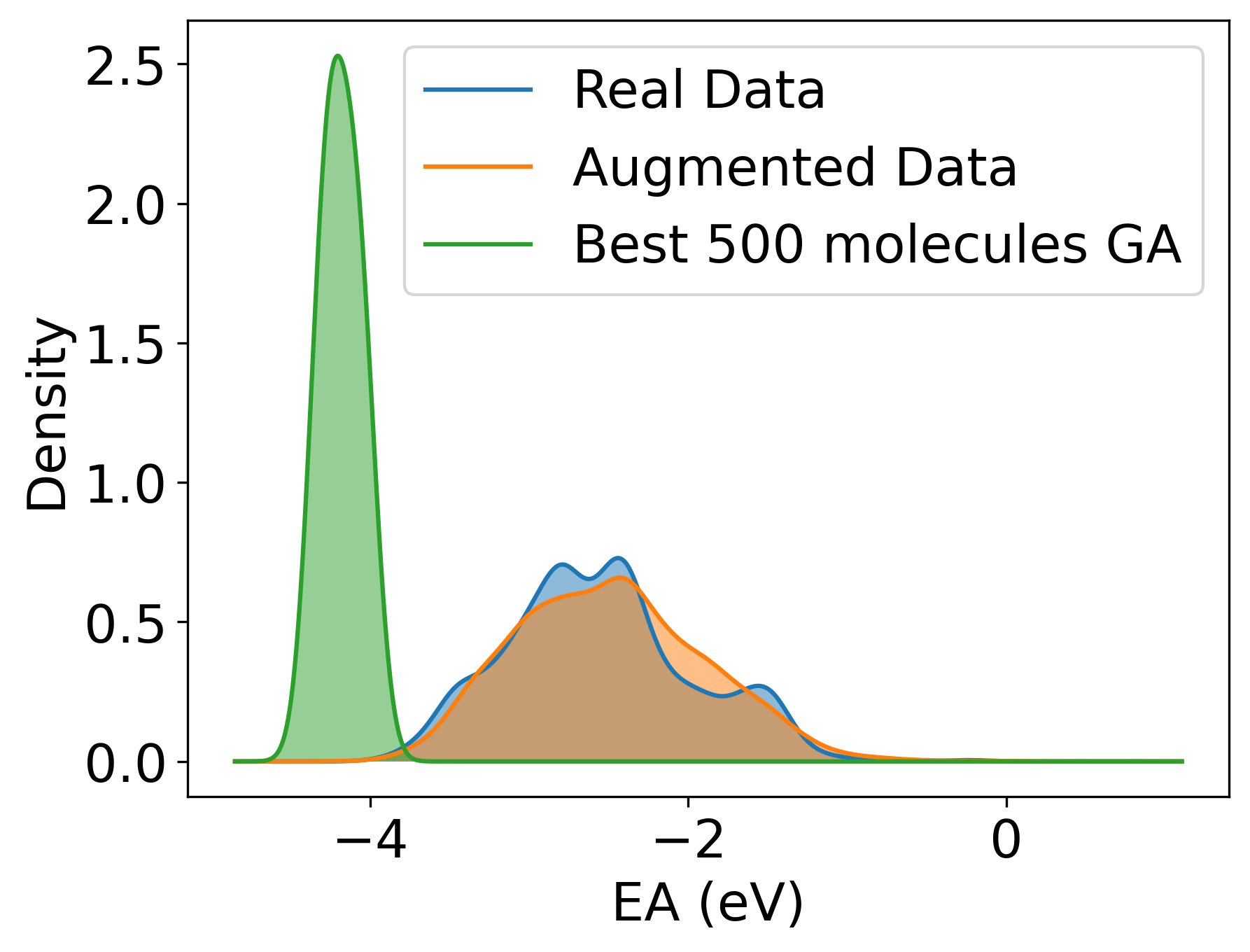}
        \caption{Electron affinity}
        \label{fig:kde_GA_EA}
    \end{subfigure}
    \begin{subfigure}[b]{0.45\textwidth}
        \centering
        \includegraphics[width=\textwidth]{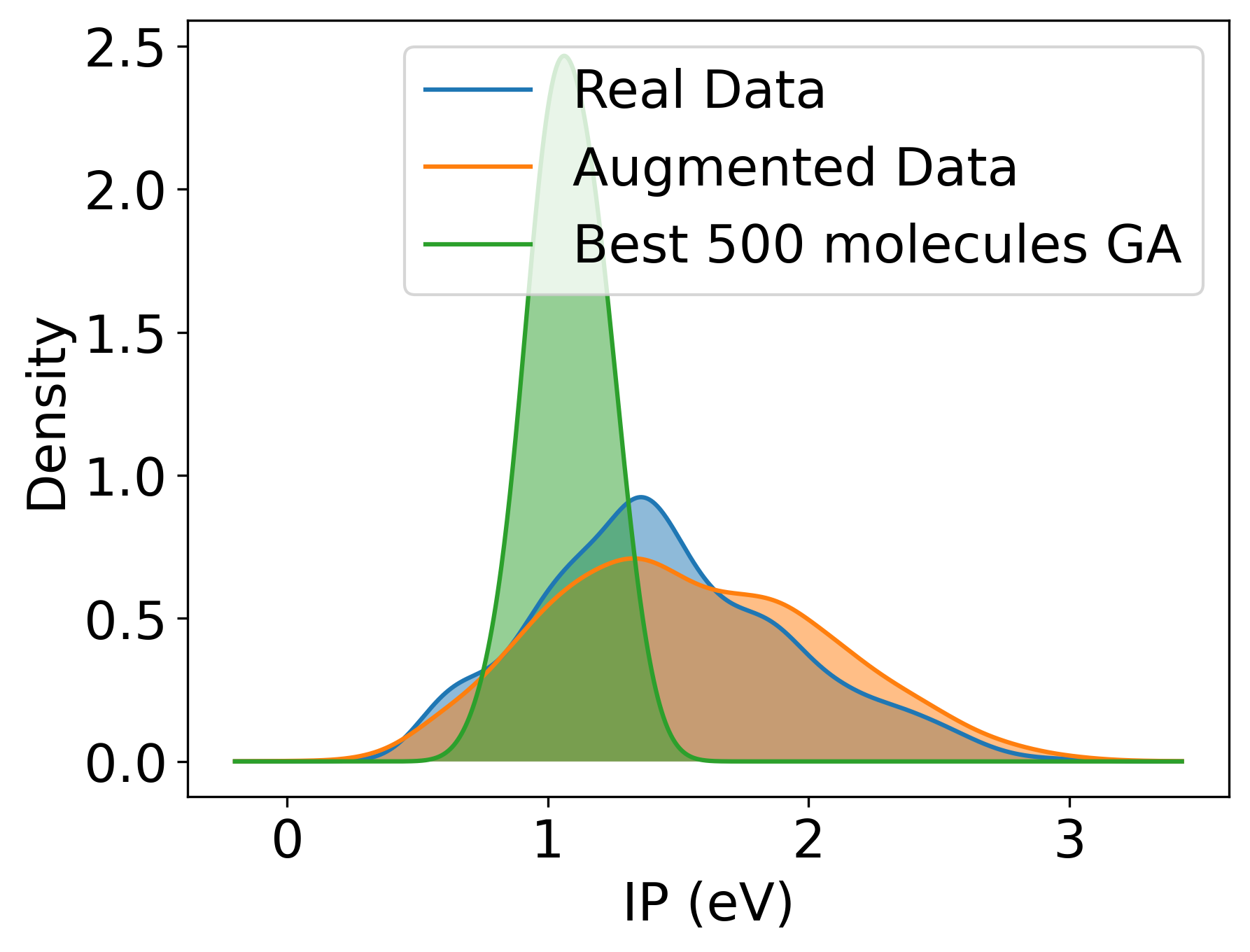}
        \caption{Ionization potential}
        \label{fig:kde_GA_IP}
    \end{subfigure}
   \caption{Kernel density estimation of property distribution in the real dataset, the augmented dataset and the best 500 molecules sampled during the inverse design using GA optimization. The plots show that optimization with a GA enables a distribution shift, generating polymers with targeted properties.}
    \label{fig:kde_GA_properties}
\end{figure}

With the objective function defined in Equation~\eqref{eq:propOpt}, we can directly apply optimization techniques to decode novel polymer photocatalysts that best align with our targets.
We compare Bayesian optimization (BO) and genetic algorithms (GA), as explained in Section~\ref{sec:propOptMethods}. Table~\ref{tab:InverseDesignResults} shows a comparison of the best objective values of polymers in the training dataset with the decoded polymers using BO and GA. Notably, both BO and GA produce top 3 candidates that better satisfy the objective than the best candidate from the dataset. Inspecting the average objective of the top 10 polymers in Table~\ref{tab:InverseDesignResults} shows that both optimization strategies outperform selecting the ten best candidates from the dataset, while we observe that the GA outperforms the BO.
One possible explanation is the computational efficiency of the GA which allows it to evaluate a greater number of molecules than the BO within the same runtime, giving it an advantage in identifying more candidates with high objective values. Thus, we conclude that with the aim to obtain a large variety of good candidate polymers, the GA, in combination with our model, is the most effective approach for inverse design. \\
The result is a library of polymer structures that adhere to the specified target objectives. Figures~\ref{fig:kde_GA_EA} and ~\ref{fig:kde_GA_IP} show the effective distribution shift towards polymers with targeted EA and IP, compared to the wide distribution of properties in the training data.

\begin{table}
  \caption{Comparison of polymers from the dataset and the generated data after optimization in the latent space. The table shows the average objective values according to Equation~\eqref{eq:propOpt} of the ten best polymers and the objective values of the top three polymers.}
  \label{tab:InverseDesignResults}
  \centering
  \begin{tabular}{lcc}
    \toprule
    Data     & $f(\vect{z})$ (Top 3)    & $f(\vect{z})$ (Average (top 10)) \\
    \midrule
    Training data  & -4.027, -4.009, -3.936 & -3.9362   \\
    Generated data (BO) &-4.451, -4.294, -4.199 & -4.0151\\
    Generated data (GA) & -4.297, -4.224, -4.160 & \textbf{-4.0705}\\
    \bottomrule
  \end{tabular}
\end{table}

\begin{figure}
    \centering
    \includegraphics[width=0.8\textwidth]{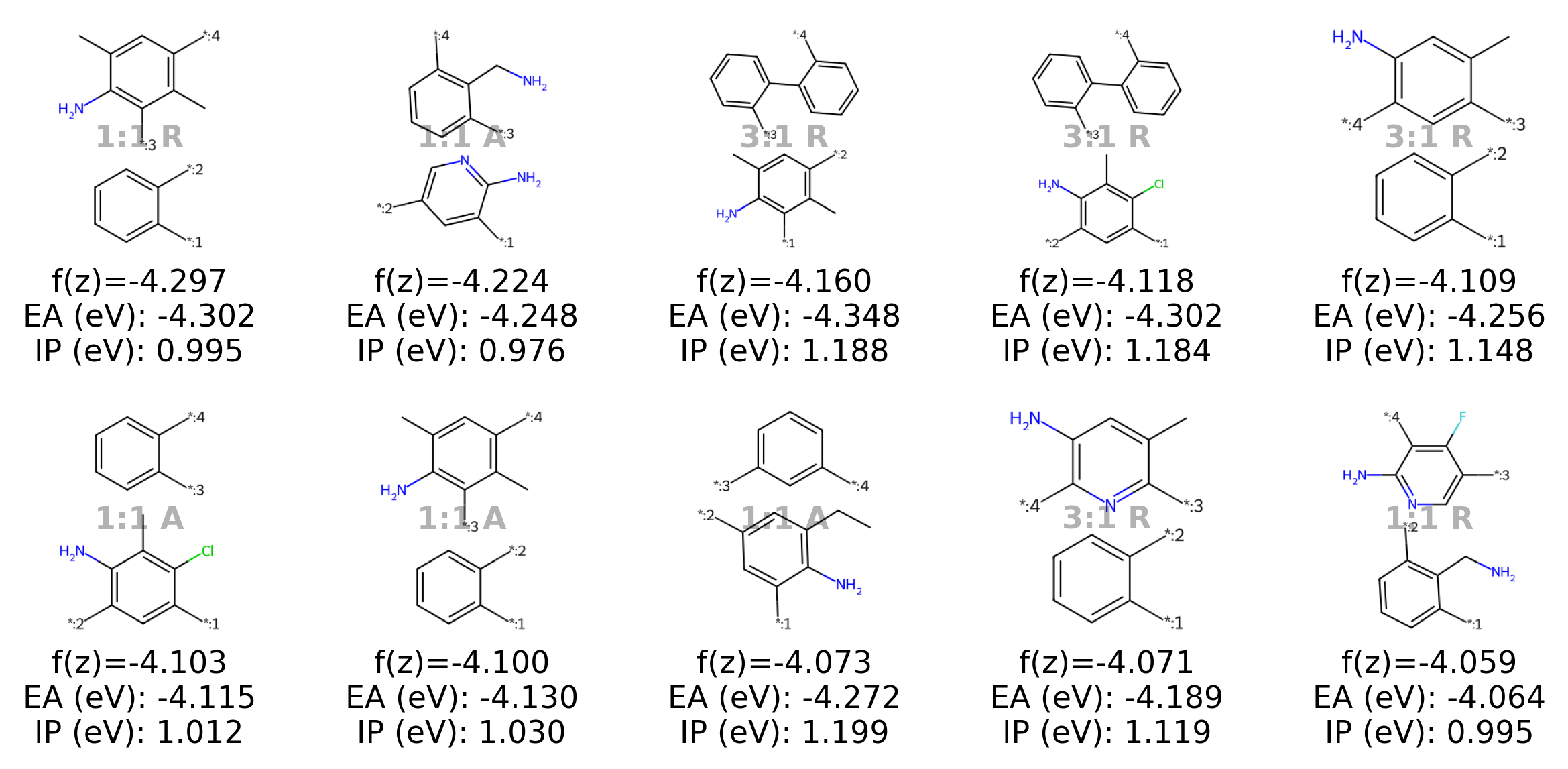}
    \caption{The top ten decoded candidates after optimization of the objective function in Equation~\eqref{eq:propOpt} in latent space using a genetic algorithm. }
    \label{fig:InvDesign_GA}
\end{figure}

\begin{figure}
    \centering
    \includegraphics[width=0.8\textwidth]{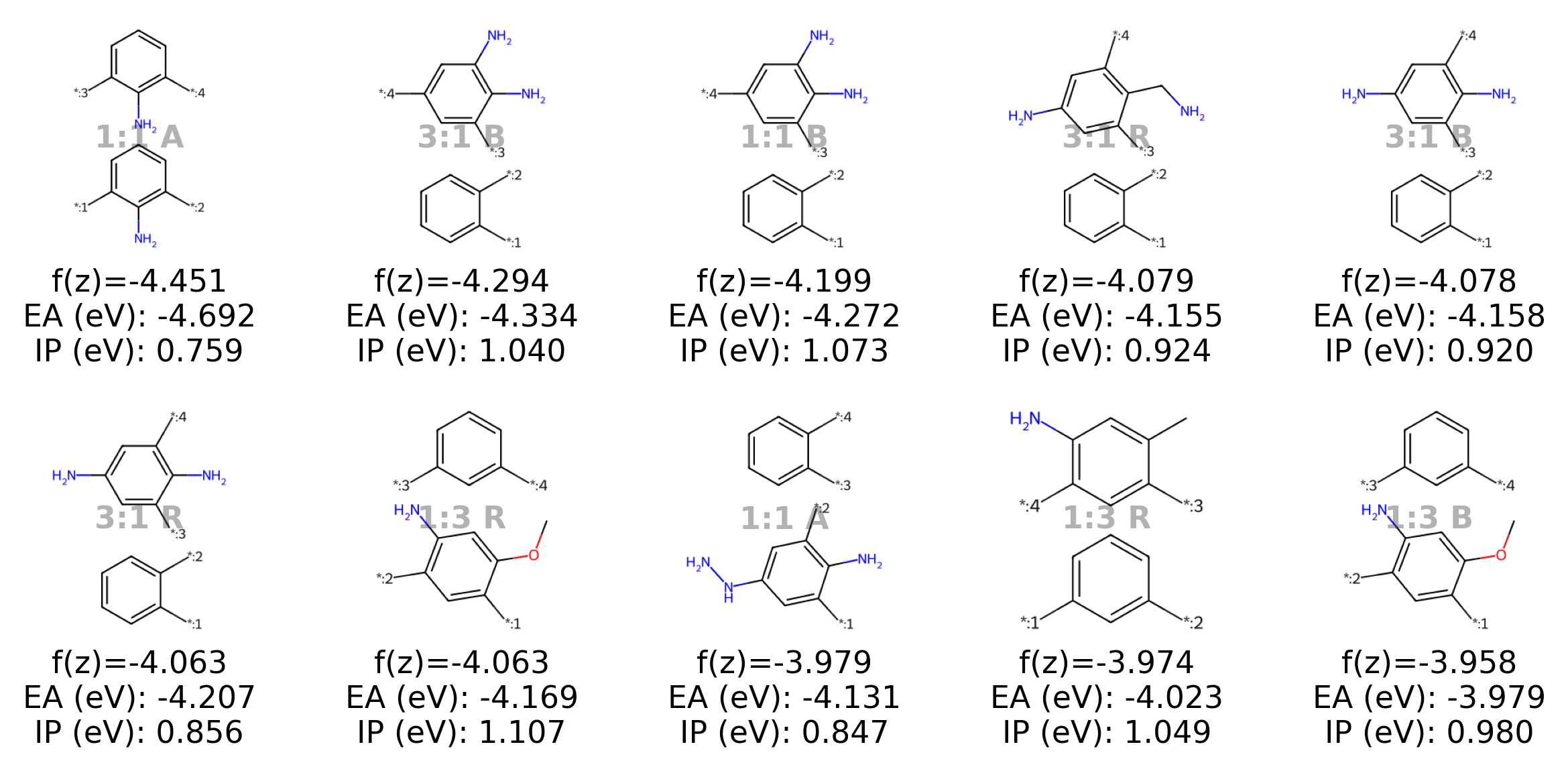}
    \caption{The top ten decoded candidates after optimization of the objective function in Equation~\eqref{eq:propOpt} in latent space using Bayesian optimization. }
    \label{fig:InvDesign_BO}
\end{figure}

As mentioned in~\citep{bai2019accelerated}, even optimal values of IP and EA according to the objective function do not guarantee high HERs. 
Note that it is important to consider additional factors (beyond electronic properties) that influence the HER for experimental validation. However, the set of candidates after the optimization likely comprises more high-performing materials than a broader set of structures and is thus very valuable for material development. Furthermore, researchers using these methods should carefully consider the importance of different properties (weights in objective function). In our setting with equal weights for both objectives, we observe that the optimization favors low EA values while sacrificing larger deviations of IP from 1~eV.


\paragraph{Discussion of the top 10 candidates}
Analysis of the best polymer candidates in Figures~\ref{fig:InvDesign_GA} and~\ref{fig:InvDesign_BO} suggests that the properties EA and IP, are primarily influenced by the monomer chemistries. Both optimization algorithms converge to a specific region in the latent space characterized by monomers of similar sizes, typically featuring one aromatic ring and at least one nitrogen atom, with rare occurences of other heteroatoms such as oxygen, fluorine and chlorine. Sulfur and bromine, which are present in the dataset, do not appear in the optimal polymers identified by our model. The chemical region that the optimization algorithms converge to is strongly influenced by the property targets defined in the objective function, which can be easily adjusted to investigate different scenarios (see Appendix~\ref{app:differentObjFuns}).

Stoichiometry and chain architecture play a less significant role. For instance, candidates two and three of the BO consist of the same monomers and block chain architecture and have very similar properties, despite the different stoichiometry. Similarly, candidate 5 and 6 only differ in the chain architecture but possess very similar properties. 
The importance of monomer chemistry compared to stoichiometry and chain architecture aligns with our observations of the latent space structure (Figures~\ref{fig:LSAMon} and~\ref{fig:LatentSpaceplotsStoichCon}) and the analysis in prior studies~\citep{aldeghi2022graph}. Nonetheless, it's important to acknowledge that for other datasets or properties, stoichiometry and chain architecture might play a more critical role in the optimization process.

\subsection{Limitations and future work}\label{sec:discussion}
Polymer properties are influenced not only by the repeat unit chemistry, but also by their higher-order structure and processing conditions. While our work incorporates stochasticity, stoichiometry of monomers and chain architectures, further factors such as weight distribution, linking structure, and processing conditions, also play crucial roles. We see the inclusion of hierarchical information in datasets and in informative representations as one of the main challenges of polymer informatics. 
We expect that this influences the design of future ML model architectures as well as the granularity of predictions.

While the dataset we use enables exploration of novel polymer photocatalysts, our results also reveal the need for greater dataset diversity in monomer chemistries and higher-order structural information (stoichiometry and chain architecture). Expanding dataset diversity across all structural levels will help prevent overfitting to monomers or e.g., a set of stoichiometries, and therefore constructing a truly continuous chemical latent space. This would naturally increase the novelty and diversity of the generated structures.
Synthesizability is another critical consideration in polymer design, yet existing metrics like the synthetic accessibility (SA) score typically encountered in small molecule discovery are limited in their application to polymers. Polymers possess a hierarchical structure beyond the monomer and their synthesis involves complex steps. Hence, developing a tailored metric that considers polymerization methods, monomer reactivity, and monomer availability is essential for translating computational predictions into experimentally realizable materials.

Lastly, while VAEs are effective at generating data within the training data distribution, they struggle with creating truly novel, out-of-distribution molecules. Techniques like reinforcement learning or genetic algorithms iteratively modifying the molecular structure (e.g. molecular graph) may be suited for out-of-distribution design but require reliable property predictors or experimental validations. When the generated molecules deviate significantly from the training data, the accuracy of property predictors is likely to diminish, underscoring the importance of robust validation methods.

\section{Conclusion}
In conclusion, our model explores a novel VAE architecture encoding polymer graphs using a graph neural network and decoding polymer strings with a Transformer. With the consideration of stochasticity of monomer ensembles including their stoichiometry and chain architecture our model represents a significant step forward in the inverse design of synthetic polymers, going beyond the repeat unit structure. We leverage a semi-supervised setup to handle partly labelled data, which is promising for a domain like polymer informatics with limited labelled data. Our model is designed to work effectively across a wide range of polymer datasets that consider the same structural levels. Additionally, we consider it to be easily adaptable, making it a useful starting point for further development as new datasets with more complex structural information emerge.\\
Finally, for the use case of conjugated copolymer-photocatalysts for hydrogen production, we demonstrate the capability of our model to generate conjugated copolymers with tailored electronic properties. We do this by using optimization in the latent space that encodes monomer combinations including monomer stoichiometries and chain architectures. Notably, the inverse design approach identifies novel copolymers that exhibit properties better than the best properties in the used training dataset. The results hold promise for accelerating the discovery of new high-performing polymer materials considering their hierarchical structure.


\section{Data availability}
The data and code can be found openly accessible on https://github.com/Intelligent-molecular-systems/Inverse\_copolymer\_design.
\section{Author contributions}
Conceptualization was carried out by J.M.W. and G.V. G.V. was responsible for data curation, formal analysis, investigation, methodology, software development, validation, and visualization. G.V. wrote the original draft of the manuscript. J.M.W. provided supervision and contributed to the review and editing of the manuscript.

\bibliographystyle{unsrtnat}
\bibliography{references}  
\newpage
\appendix
\section{Additional details on methods}
\subsection{Polymer string tokenization}\label{app:tokenization}
For the tokenization of the polymer strings we implement a tokenizer that is inspired by the Regression Transformer~\cite{born2022RT} (RT) tokenization. The strings are tokenized using a combination of a SMILES tokenization and floating point number tokenization. The SMILES tokenization alone uses the same vocabulary for the digits in the SMILES string and the digits in the floating point numbers. The RT-tokenization distinguishes digits in SMILES and the digits in floating point numbers. We demonstrate the difference using the number 0.125 and the monomer string [*:1]c1cc2sc3cc([*:2])sc3c2s1. In red we highlighted the tokens that are encoded with the same vocabulary but represent a different meaning. 
\begin{itemize}
       \item[]SMILES-tokenization: 0.125 $\rightarrow$ 0 . \textcolor{red}{1} \textcolor{red}{2} 5
       \item[]SMILES-tokenization:  [*:1]c1cc2sc3cc([*:2])sc3c2s1 $\rightarrow$ [* :1 ] c \textcolor{red}{1} c c \textcolor{red}{2} s c 3 c c ( [* :2 ] ) s c 3 c \textcolor{red}{2} s \textcolor{red}{1}
\end{itemize}
Using the RT-tokenization, the digits in floating point numbers are enriched by the information of their decimal position (..,0,-1,-2, -3, ...) which mitigates this issue of using the same vocabulary: 
\begin{itemize}
       \item[]RT-tokenization: 0.125 $\rightarrow$ 0$\_$0 . \textcolor{blue}{1$\_$-1} \textcolor{blue}{2$\_$-2} 5$\_$-3
       \item[]RT-tokenization:  [*:1]c1cc2sc3cc([*:2])sc3c2s1 $\rightarrow$ [* :1 ] c \textcolor{orange}{1} c c \textcolor{orange}{2} s c 3 c c ( [* :2 ] ) s c 3 c \textcolor{orange}{2} s \textcolor{orange}{1}
\end{itemize}
Further the decimal tokenization could be used in future work together with numerical encodings as demonstrated in~\citep{born2022RT}.

\subsection{Variational lower bound of graph-2-string VAE}\label{app:lossderivation}
A VAE is trained by maximizing the Variational Lower Bound, which is defined as follows for one datapoint $\vect{x}^{(i)}$
\begin{equation}\label{eq:VaLoBo}
    \mathcal{L}(\vect{\theta}, \vect{\phi}; \vect{x}^{(i)}) = \mathbb{E}_{q_{\vect{\phi}}(\vect{z}|\vect{x}^{(i)})} \left[ \log p_{\vect{\theta}}(\vect{x}^{(i)}|\vect{z}) \right] -D_{KL}(q_{\vect{\phi}}(\vect{z}|\vect{x}^{(i)}) || p_{\vect{\theta}}(\vect{z})) 
\end{equation}
balancing the maximization of the reconstruction term $\mathbb{E}_{q_{\vect{\phi}}(\vect{z}|\vect{x}^{(i)})}[\log p_{\vect{\theta}}(\vect{x}^{(i)}|\vect{z})]$ with minimization of the regularization term $D_{KL}(q_{\vect{\phi}}(\vect{z}|\vect{x}^{(i)})||p(\vect{z}))$, where the prior $p(\vect{z})$ is a normal distribution $\mathcal{N}(0, I)$. During training, the negative of the lower bound is minimized, as described in \citep{kingma2013VAE}.

As a result of the modified components of the VAE, encoding a graph and decoding a string, the variational lower bound in Equation~\ref{eq:VaLoBo} can be rewritten as 
\begin{equation}\label{eq:ElboG2S}
    \mathcal{L}(\vect{\theta}, \vect{\phi}; \vect{x_S}^{(i)}, G^{(i)}) = \underbrace{\mathbb{E}_{q_{\vect{\phi}}(\vect{z}|G^{(i)})} \left[ \log p_{\vect{\theta}}(\vect{x_S}^{(i)}|\vect{z}, G^{(i)}) \right]}_{-\mathcal{L}_{rec}} -\underbrace{D_{KL}(q_{\vect{\phi}}(\vect{z}|G^{(i)}) || p_{\vect{\theta}}(\vect{z}))}_{\mathcal{L}_{KLD}}.
\end{equation}
The loss that is minimized during training is the negative of Equation~\eqref{eq:ElboG2S} with the addition of a hyperparameter $\beta$ to balance the two loss terms. 
\begin{equation}\label{eq:LossG2S_app}
    \mathcal{L}=\mathcal{L}_{Rec} + \beta \cdot\mathcal{L}_{KLD}
\end{equation}

\subsection{Choice of target values in objective function}\label{app:choiceTargetVals}
The goal in our study is to generate novel polymers with targeted property values indicating a high HER. To do so, we leverage the continuous latent space of our model, allowing for optimization of the latent variables to decode candidates with desired properties. In our case study the ideal property to optimize would be the HER, which is not given for the dataset we used to train our model. However, \citet{bai2019accelerated} found that materials with more negative electron affinity (EA) and more positive ioinization potential (IP) showed better HERs. This is likely related to the optical gap, approximated as |EA - IP|; larger optical gaps often correlate with higher HER. Thus, maximizing the optical gap by minimizing EA and maximizing IP could help identify high-performance materials. Looking at the study's results more closely, it showed that the HER was nearly zero for polymers with positive EA values on the standard hydrogen electrode (SHE) scale and peaked at an EA of around -2\,eV. This peak was at the lower end of the EAs observed, suggesting optimal driving force for the proton reduction when minimizing this property. The number of materials with high observed HERs increased with rising IP, peaking around 1\,eV before declining. This indicates the importance of balancing the driving force for oxidation. We conclude that, to optimize copolymers for photocatalytic activity, one could target minimal EA values and IP values near 1\,eV (see Section~\ref{sec:propOptMethods} and the objective function in Equation~\eqref{eq:propOpt}.   

\section{Additional results}
\subsection{Impact of higher-order structure on polymer properties}
\begin{figure}[H]
    \centering
    \includegraphics[width=0.7\textwidth]{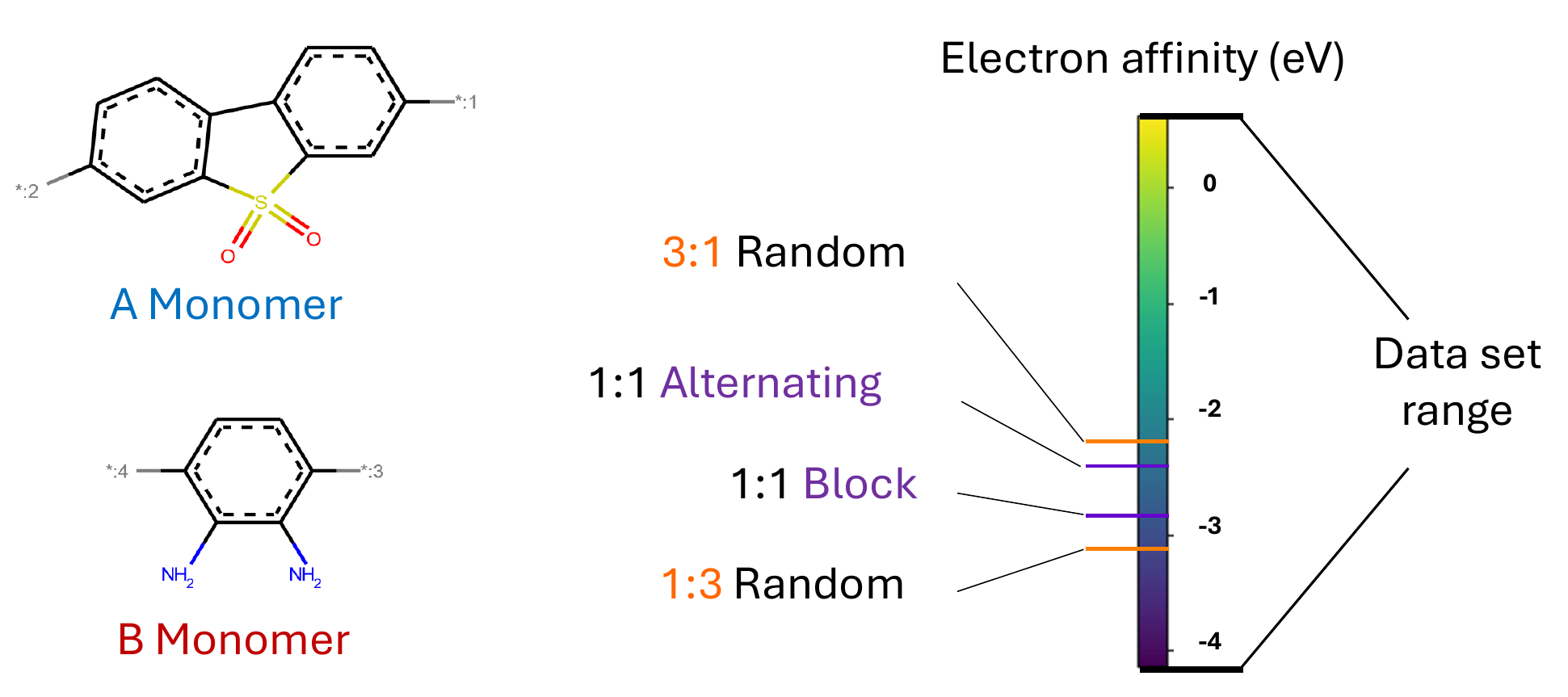}
    \caption{For a fixed monomer combination (left), changing the stoichiometry and chain architecture influences the electron affinity of the polymer (right). }
    \label{fig:property_impact_HOS}
\end{figure}

Figure~\ref{fig:property_impact_HOS} shows an example monomer combination found in the dataset. Changing the stoichiometry from 3:1 to 1:3 leads to a significant drop of the EA compared to the data set range. Further, for a fixed stoichiometry, changing the chain architecture from alternating to block structure leads to a non negligible drop in the EA of the polymer material. This example illustrates that the higher order structure matters to learn accurate structure-property relationships.

\subsection{Polymer property prediction performance}\label{app:propPred}
\begin{figure}[H]
    \centering
    \begin{subfigure}[b]{0.45\textwidth}
        \centering
        \includegraphics[width=0.9\textwidth]{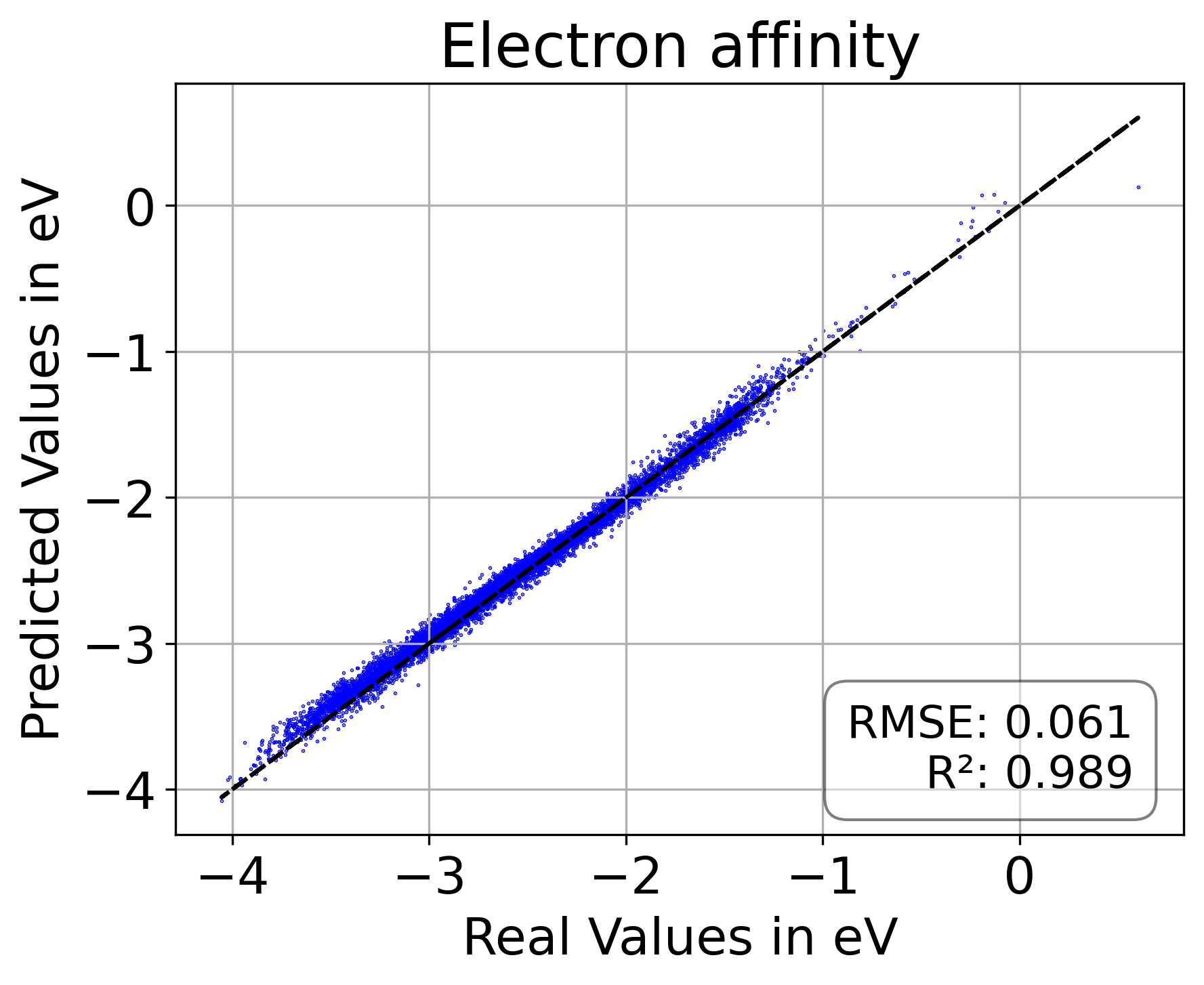}
        \caption{Property prediction accuracy of EA}
        \label{fig:PPperf_EA}
    \end{subfigure}
    \begin{subfigure}[b]{0.45\textwidth}
        \centering
        \includegraphics[width=0.9\textwidth]{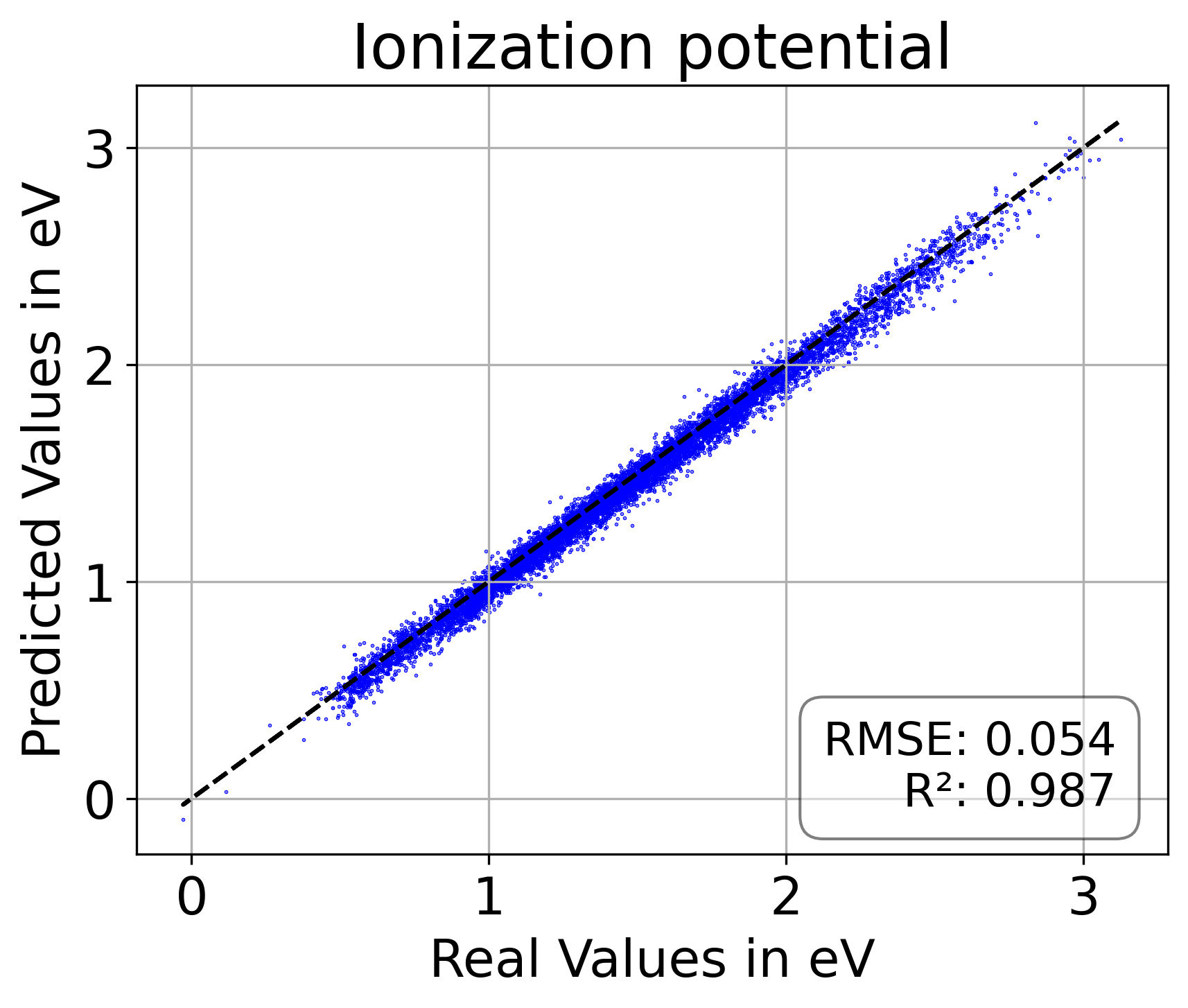}
        \caption{Property prediction accuracy of IP}
        \label{fig:PPperf_IP}
    \end{subfigure}
    \caption{Evaluation of the property prediction performance for the held out testset. }
    \label{fig:Property_prediction_performance}
\end{figure}

To assess how well properties are predicted for unseen data, we encode the test set to the latent representations and use the trained neural network on top of the latent space to predict the properties. Figure~\ref{fig:Property_prediction_performance} shows a high performance, comparable to the values reported in~\citep{aldeghi2022graph}. This is important to ensure accurate property values as feedback for the optimization algorithms in the inverse design approach.

\subsection{Additional latent space plots: Stoichiometry and chain architecture}
\begin{figure}[H]
    \centering
    \begin{subfigure}[b]{0.45\textwidth}
        \centering
        \includegraphics[width=\textwidth]{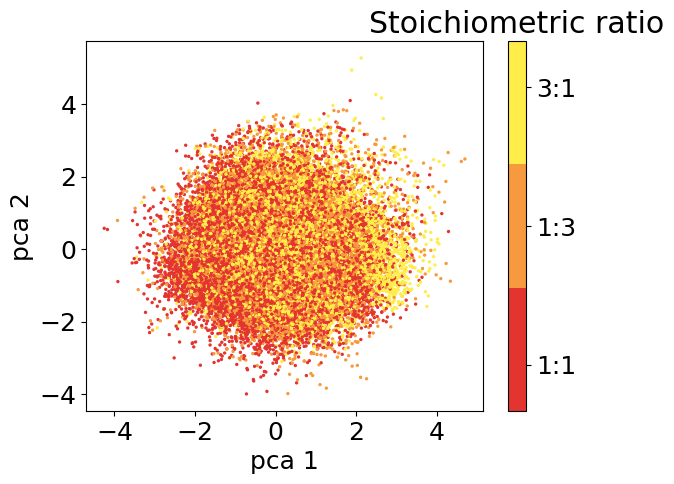}
        \caption{Colored by stoichiometry}
        \label{fig:LSstoich}
    \end{subfigure}
    \begin{subfigure}[b]{0.45\textwidth}
        \centering
        \includegraphics[width=\textwidth]{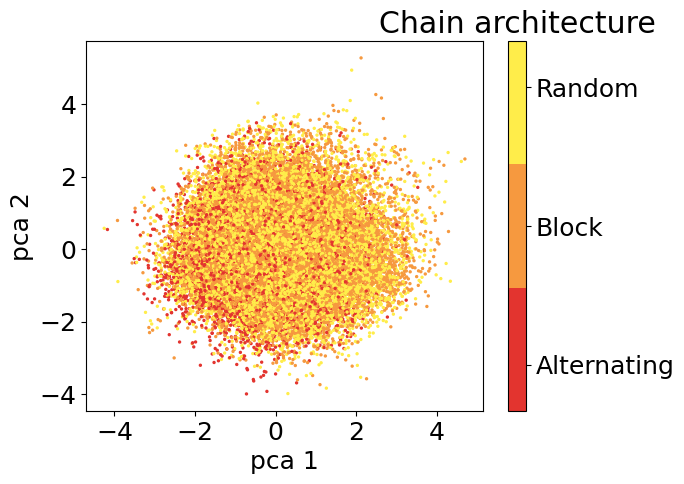}
        \caption{Colored by chain architecture}
        \label{fig:LScon}
    \end{subfigure}
    \caption{Visualization of the two first principal components of the latent space of the training data. Coloring by stoichiometry and chain architecture only shows local organization according to these structural characteristics of the polymer.}
    \label{fig:LatentSpaceplotsStoichCon}
\end{figure}

\subsection{Polymer design without optimization}\label{app:DesignWOPPO}
This analysis focuses on generating polymers that are similar to the best performing photocatalyst in \citep{bai2019accelerated}.
We make use of two common techniques. First, we can sample the latent neighborhood of good polymer candidates that showed high HERs in literature. Second, we can perform interpolation in latent space between interesting candidates. Given our well structured latent space, as demonstrated in Section~\ref{sec:LSstructure}, these approaches should help to yield a higher fraction of high-performing candidate polymers than just randomly sampled ones. \\

\subsubsection{Sampling the neighborhood of high-performing polymer}\label{app:samplingaroundbest} 
\begin{figure}
    \centering
    \input{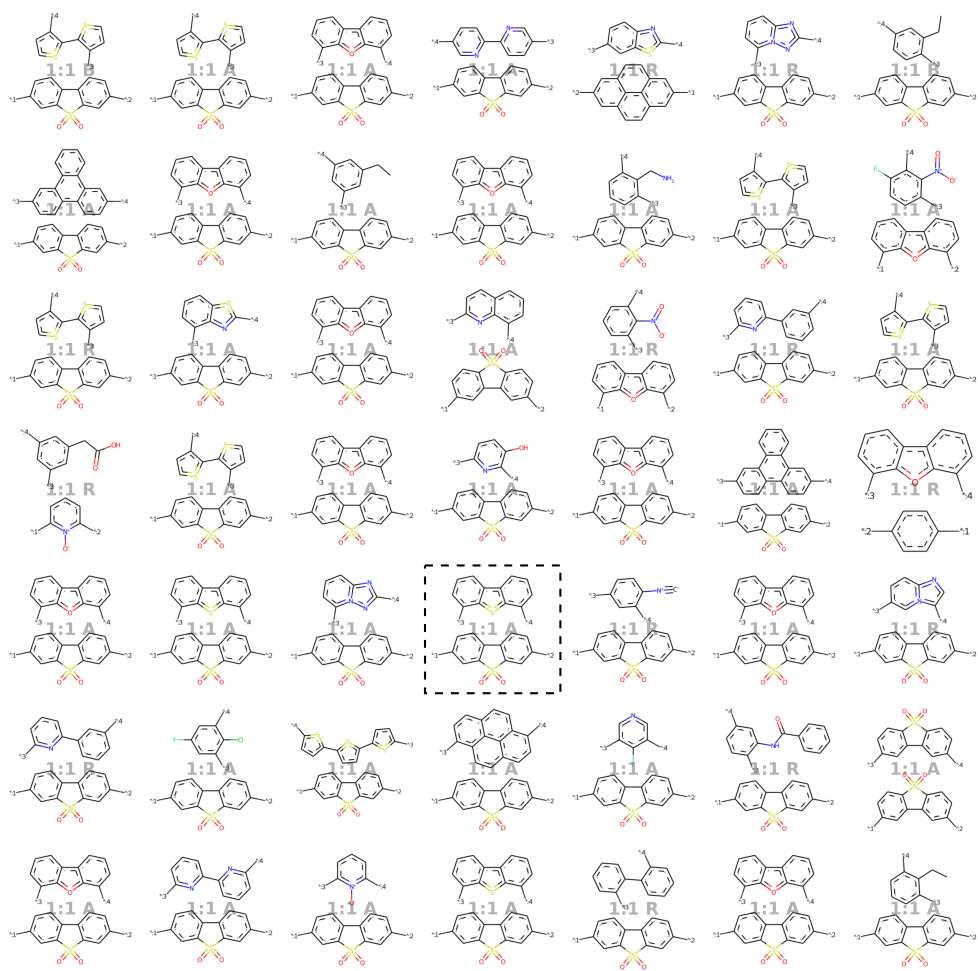} 
    \caption{Sampling around the best experimental photocatalyst found in~\cite{bai2019accelerated}. The best photocatalyst is first encoded to the latent space. By adding small bits of random noise to the latent representation, we obtain new latent vectors that can be decoded to novel similar polymer photocatalysts. }
    \label{fig:moleculesSeed}
\end{figure}

Sampling around polymers that are known to be high-performing photocatalysts can prove as a powerful approach to generate novel polymers with similarly good or better performance. To do so, we take one high-performing photocatalysts as a starting point and encode the respective graph to a continuous latent vector $\vect{z}_{seed}$. Next, we repeatedly add small noise (drawn from $\mathcal{N}(0, 0.25)$) to obtain new latent vectors. We can then decode these new latent vectors to structures. 
Figure~\ref{fig:moleculesSeed} shows the results of sampling around the polymer with the best experimental HER in \citet{bai2019accelerated}'s study, highlighted in black dashed lines. The novel molecules we obtained by sampling around the latent vector of the encoded seed molecule possess a variety of changes from the encoded polymer, where monomer B is varied most. We see that monomer A is altered less, attributable to the limited monomer A variety in the dataset. Same holds for the stoichiometry while the chain architecture is varied more than stoichiometry, meaning that it has less strong signal in the latent space than the stoichiometry (less robust to changes in the latent vector). 

\subsubsection{Interpolation between interesting candidates}
\begin{figure}
    \centering
    \includegraphics[width=0.5\textwidth]{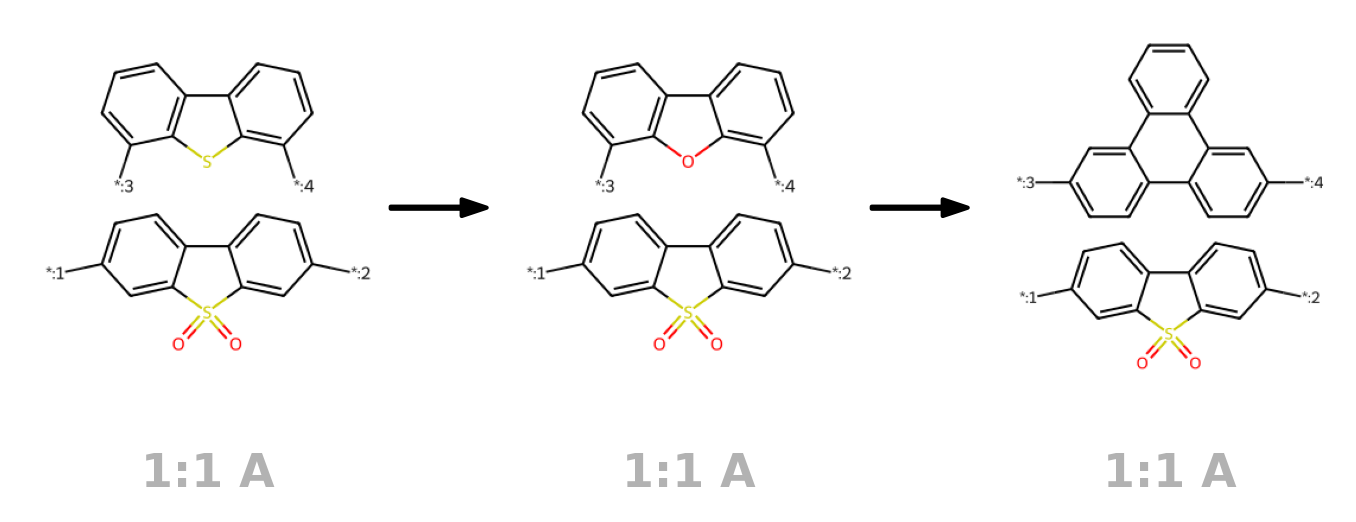}
    \caption{Interpolation in the latent space between the two best copolymer photocatalysts found in \cite{bai2019accelerated}. Starting from the best candidate polymer on the left, we observe only one polymer on the interpolation path to the second best candidate.}
    \label{fig:Interpolation_twoBest}
\end{figure}

\begin{figure}
    \centering
    \includegraphics[width=\textwidth]{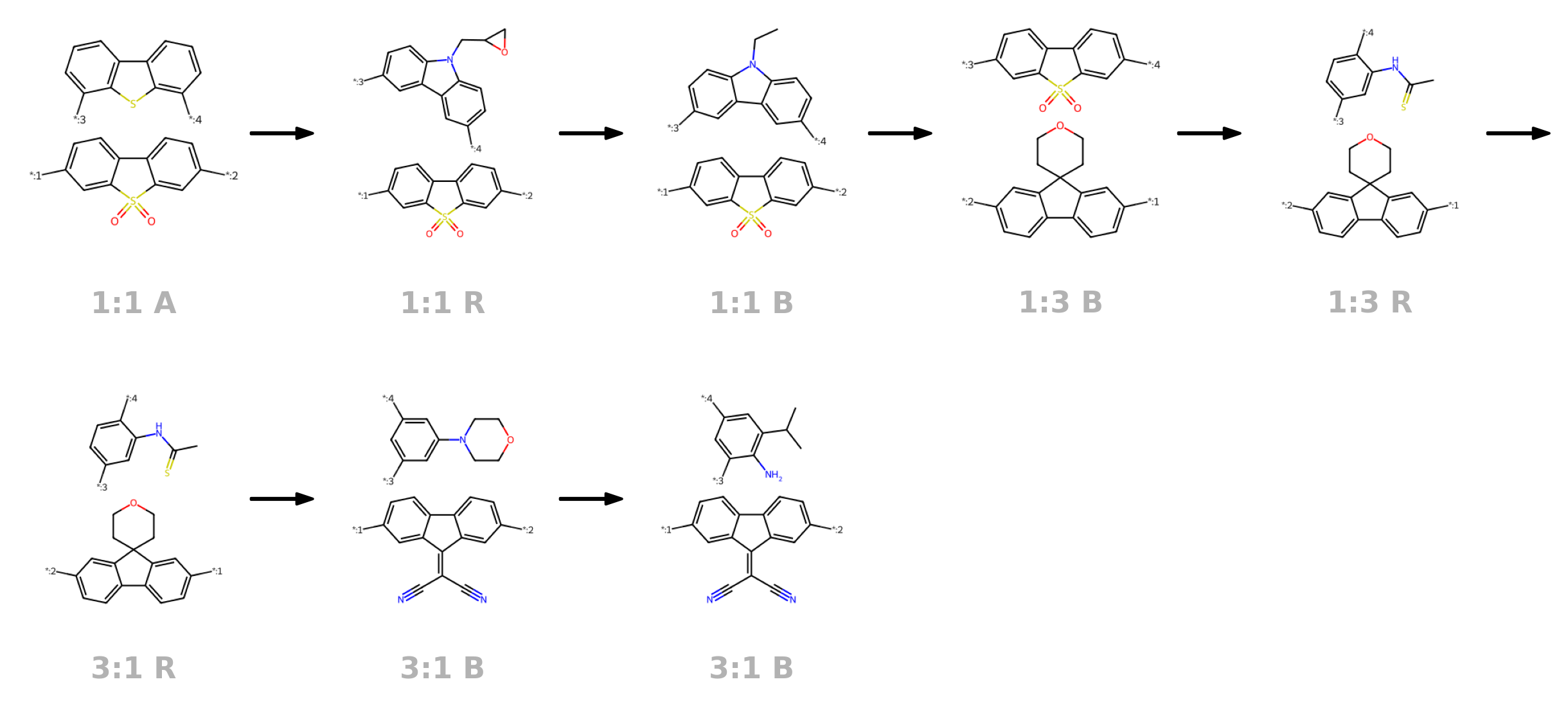}
    \caption{Interpolation in the latent space between the best copolymer photocatalysts found in \cite{bai2019accelerated} and one random polymer. We observe a step-wise change from on the monomer but level but also on the higher order structural levels, i.e. the monomer stoichiometry and chain architecture.}
    \label{fig:Interpolation_BestRand}
\end{figure}

The second strategy to sample novel polymers without property optimization is interpolation in the latent space. With the aim to find promising novel photocatalysts we can for instance interpolate between the two best polymers found in \citep{bai2019accelerated}. For this we assume that both polymers possess a 1:1 monomer stoichiometry and alternating chain architecture. However, the two best candidates are structurally very similar: Monomer A is the same and monomer B is only slightly different. As a result, the interpolation only leads to one additional candidate that is found on the interpolation path (see Figure~\ref{fig:Interpolation_twoBest}). While we expect the polymers to be close in the latent space, we could think of more candidates that lie on the interpolation path. 
Generally, interpolation is more interesting in scenarios where we want to know the path in the chemical space between two structurally substantially different molecules. For instance, we can interpolate between the best molecule and a random molecule in the dataset, leading to a more interesting path as visualized in Figure~\ref{fig:Interpolation_BestRand}. 

\subsection{Additional inverse design experiments with varied property targets}\label{app:differentObjFuns}
In this section we show two additional experiments, where we varied the objective function for the optimization with the GA. 
\subsubsection{Target electronic properties of a known high-performing photocatalyst}\label{app:BestMolDesignApproach}
\begin{figure}
    \centering
    \input{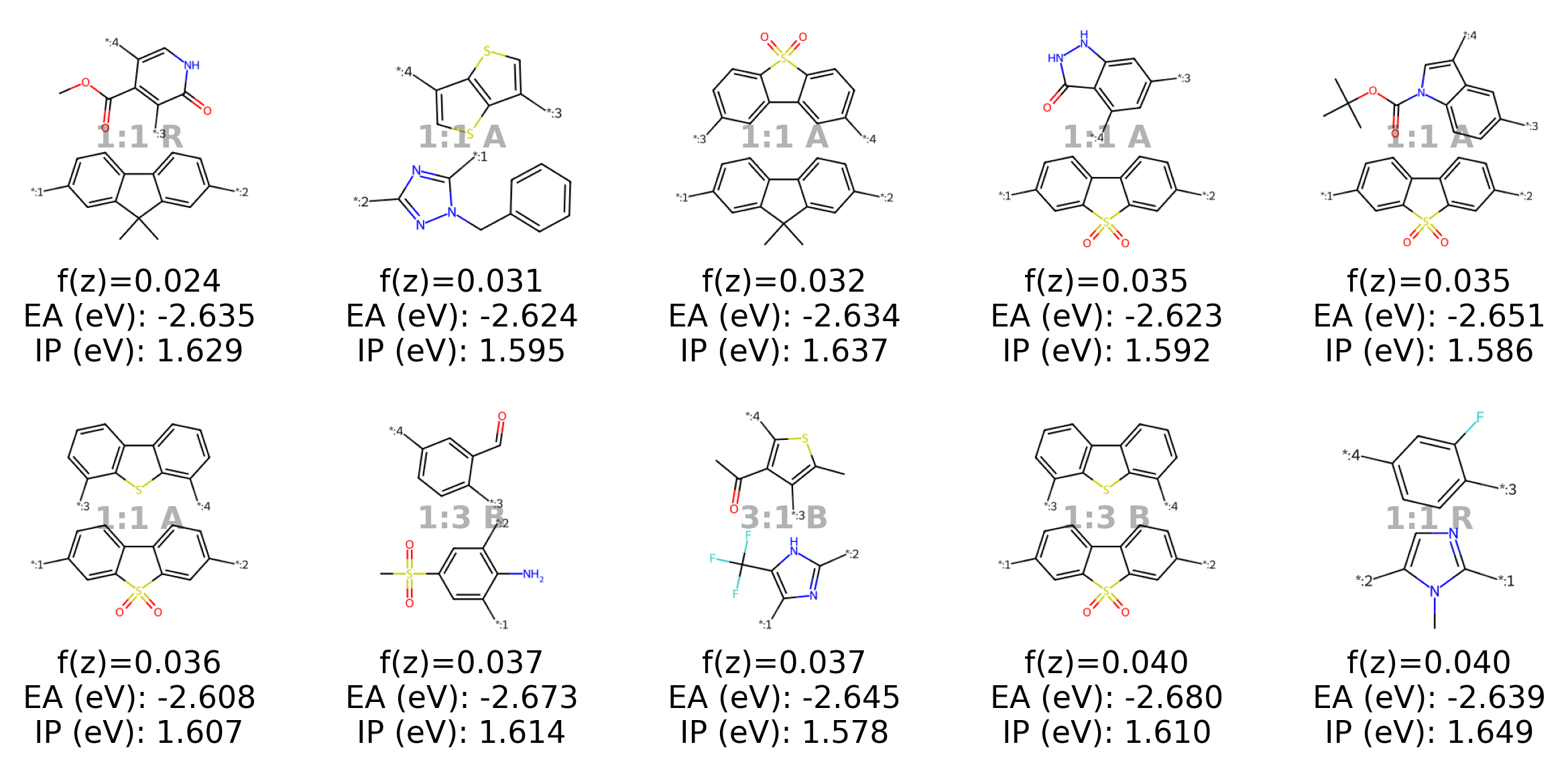} 
    \caption{Results when minimizing the objective function using a genetic algorithm. This is the modified objective function with EA and IP targeting the values of the best experimental polymer. In black dashed line we highlighted the polymer that we used to define the objectives. It is found as a solution by the GA but not as the best candidate which is contradictive. This results from small errors made by the property predictor.}
    \label{fig:InvDesign_GA_mimickbest}
\end{figure}

Let us consider the best-performing molecule in \citet{bai2019accelerated}'s study which has an EA of approximately -2.64~eV and an IP of 1.61~eV (as reported in the used training dataset~\cite{aldeghi2022graph}. These values differ markedly from the targets in the objective function in Section~\ref{sec:inversedesign}, which aimed for minimal EAs (converging to around -4~eV) and an IP of 1~eV.
Hence, we observe that the results in Figure~\ref{fig:InvDesign_GA_mimickbest} differ significantly from those in Figure~\ref{fig:InvDesign_GA}. Interestingly, the sixth-best candidate (indicated by black dashed lines) is identical to the molecule from the experimental study we used to specify the property targets. It should be the best candidate, matching the objective values exactly. However, despite the very high accuracy of the property predictor (see Section~\ref{app:propPred}), there is a small prediction error. We also see a variety of structures that satisfy the objective function well but differ structurally from the best polymer reported in the literature. 
\subsubsection{Change of target values according to dataset statistics}
Second, we adjust the target for IP and EA to the value that exhibits the largest number of well-performing (threshold defined in~\cite{bai2019accelerated}) photocatalysts from their experimental validation. As a result, we specify an EA of -2~eV and IP of 1.2~eV as optimal. 
Also for this modified objective function we observe a different region of the polymer space to be optimal compared to the results in Section~\ref{sec:inversedesign}.

An interesting detail for both of these experiments is, that the genetic algorithm (GA) converges faster for these two objective functions compared to the objective function in Equation~\eqref{eq:propOpt} and experiment in Section~\ref{sec:inversedesign}. This suggests that specifying property values as target is a simpler task than the open-ended minimization of EA. This could also be due to EAs <-4 being slightly outside of the distribution of the dataset.

\begin{figure}
    \centering
    \includegraphics[width=0.8\textwidth]{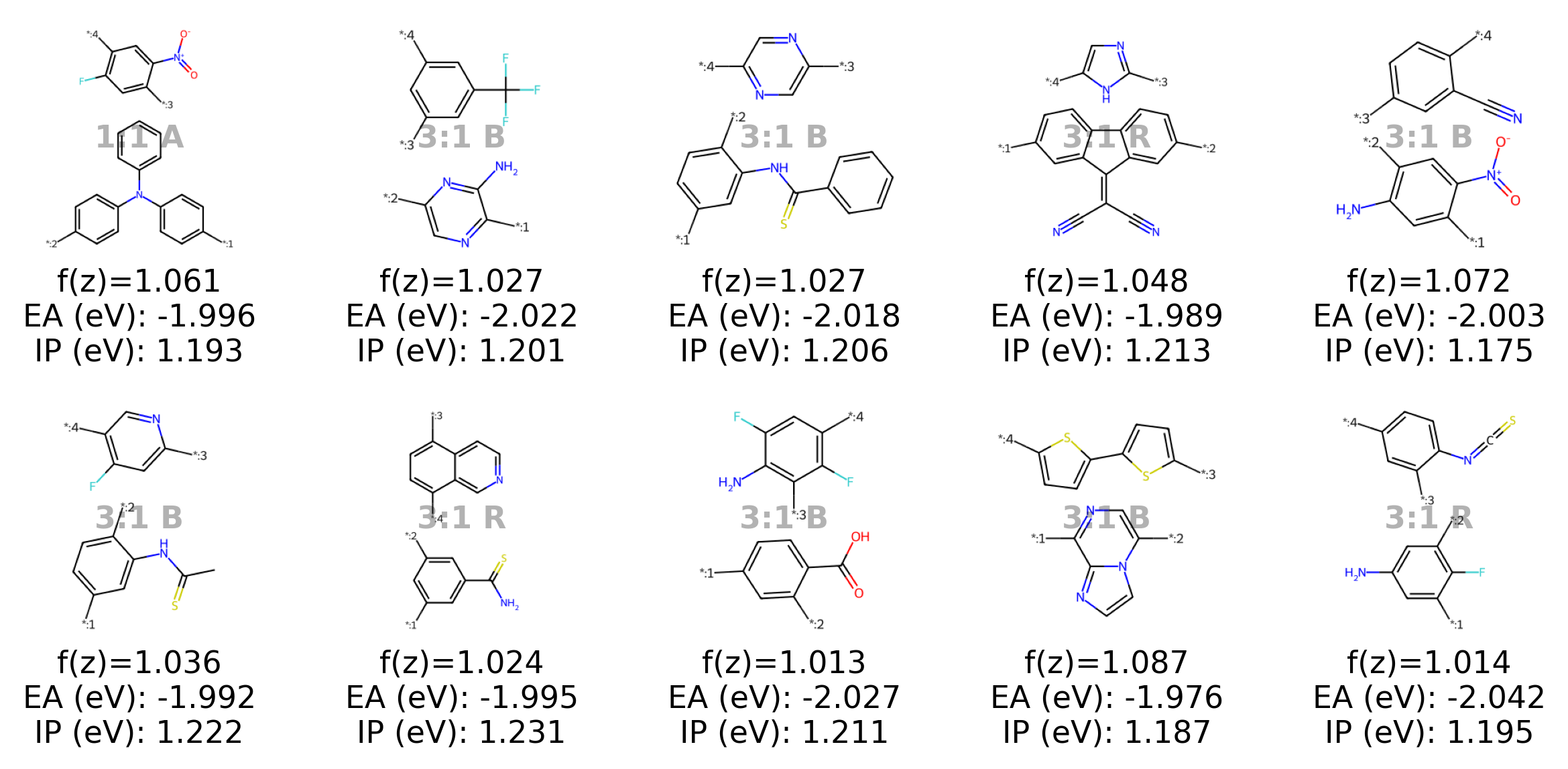}
    \caption{Results when minimizing the objective function using a genetic algorithm. This is the modified objective function targeting an EA of -2~eV and an IP of 1.2~eV. The objective matches the property values where most well-performing materials were found according to~\citep{bai2019accelerated}}
    \label{fig:InvDesign_GA_mimickpeak}
\end{figure}

\subsection{Additional smoothness plot around known high-performing photocatalyst}
\begin{figure}[H]
    \centering
    \input{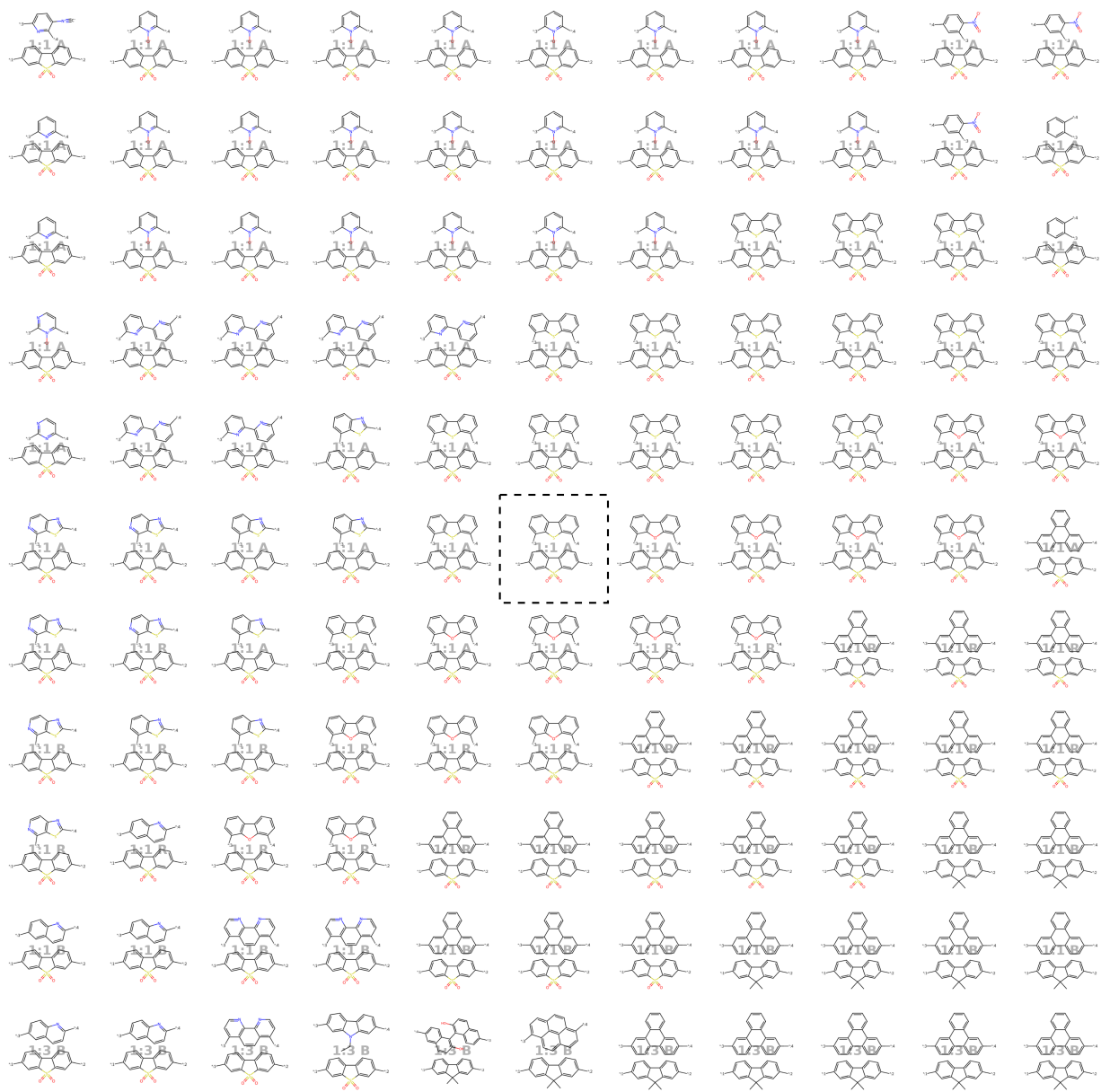} 
    \caption{Visualization of the molecular neighborhood (as proposed by \citet{kusner2017grammarVAE}) of the best performing polymer found in~\citep{bai2019accelerated}. Moving around the encoded molecule in black dashed lines in the center, we can observe step-wise changes in monomer A (bottom one), monomer B (top one), stoichioemtry and chain architecture.}
    \label{fig:moleculesNeighborhood_bestBai}
\end{figure}



\end{document}